\documentclass{aa}
\usepackage{graphicx}
\usepackage[varg]{txfonts}
\usepackage[colorlinks = true,
            linkcolor = blue,
            urlcolor  = blue,
            citecolor = blue,
            anchorcolor = blue]{hyperref}
\usepackage{color}
\usepackage[normalem]{ulem}

\newcommand{\eq}[1]{Eq. (\ref{eq:#1})}

\newcommand{\fg}[1]{Fig.\ \ref{fig:#1}}

\newcommand{\Fg}[1]{Figure\ \ref{fig:#1}}

\newcommand{\api}[1]{Appendix\ \ref{ap:#1}}

\begin{document}

   \title{The radial structure of planetary bodies formed by the streaming instability}

   \author{R.G. Visser$^{a}$\thanks{Corresponding author. E-mail:  r.g.visser@uva.nl}, J. Dr\k{a}\.{z}kowska$^b$,  C. Dominik$^a$}

   \institute{$^{a}$ Anton Pannekoek Institute for Astronomy (API), University of Amsterdam,
              Science Park 904, 1098XH, Amsterdam\\
              $^b$ University Observatory, Faculty of Physics, Ludwig-Maximilians-Universit\"at M\"unchen, Scheinerstr. 1, 81679 Munich, Germany }

   \date{Accepted on January 21, 2021}
 
\abstract{Comets and small planetesimals are believed to contain primordial building blocks in the form of millimeter to centimeter sized pebbles. One of the viable growing mechanisms to form these small bodies is through the streaming instability (SI) in which pebbles cluster and gravitationally collapse toward a planetesimal or comet in the presence of gas drag. However, most SI simulations are global and lack the resolution to follow the final collapse stage of a pebble cloud within its Hill radius. We aim to track the collapse of a gravitationally bound pebble cloud subject to mutual collisions and gas drag with the representative particle approach. We determine the radial pebble size distribution of the collapsed core and the impact of mutual pebble collisions on the pebble size distribution. We find that virial equilibrium is never reached during the cloud evolution and that, in general, pebbles with a given Stokes number (St) collapse toward an optically thick core in a sequence from aerodynamically largest (St $\sim$ 0.1) to aerodynamically smallest (St $\sim 2 \times 10^{-3}$). We show that at the location where the core becomes optically thick, the terminal velocity $v_{t,*} \sim 60 \ \mathrm{m \ s^{-1}} \mathrm{St}^2$ is well below the fragmentation threshold velocity. While collisional processing is negligible during cloud evolution, the collisions that do occur are sticking. These results support the observations that comets and small planetary bodies are composed of primordial pebbles in the millimeter to centimeter size range.}
  

\maketitle
\section{Introduction}
\label{sec:intro}
The formation mechanism of planetesimals and comets remains uncertain. While micrometer grains can grow through coagulation to roughly centimeter sized pebbles \citep{DominikTielens1997,Birnstieletal2012}, growth stagnates due to the bouncing barrier \citep{Zsometal2010} and the fragmentation barrier \citep{BlumMunch1993}. However, if growth to meter sized boulders is obtained by coagulation, these objects drift toward the star within several hundred orbital timescales due to the nebular gas \citep{Weidenschilling1977A, Nakagawa1986}. It was long thought that the increased sticking properties of water ice could circumvent the bouncing barrier \citep{Wadaetal2009,Gundlachetal2011}, though recent lab experiments have shown that this advantage only holds in a rather narrow disk temperature range \citep{MusiolikWurm2019} and that relative velocities quickly result in fragmentation\citep{Blum2008}. 

The currently favored growth model is the gravitational collapse of a pebble cloud due to highly concentrated clumps of solids \citep{JohansenEtal2007,JohansenEtal2009} induced by the streaming instability (SI) \citep{YoudinGoodman2005}. The high solid-to-gas ratios needed to trigger SI are not straightforward to achieve in typical protoplanetary disks. Studies do however show that so-called pressure bumps lead to over-densities of solids \citep{Whipple1972, Braueretal2008,Drazkowskaetal2013}. Furthermore the re-condensation of icy pebbles just outside the snowline can lead to a pileup of solids, enhancing pebble surface densities by a factor of five or more \citep{SchoonenbergOrmel2017,DrazAlibert2017}.

A limited spatial resolution in large-scale SI simulations creates difficulties in following the final collapse phase of these formed clumps. This phase is vital in understanding the final structure of the solid core and how this compares to observations of planetary bodies. N-body simulations of a collapsing spherical cloud of pebbles with significant rotation produce similar mass binaries, such as Pluto and Charon \citep{Nesvornyetal2010}. Also, recent large-scale hydrodynamical simulations match the 80 percent prograde binary rotation direction observed in KBOs \citep{Nesvornyetal2019}. 

Detailed observational data from comets obtained by the ESA Rosetta probe provide comparison material for the pebble cloud collapse model as a potential comet formation mechanism. Comets have a porosity of up to 80 percent, are several kilometers in size, have a low bulk density of roughly 0.5 g cm-3 \citep{Groussinetal2019}, and appear to have a layered pebble-pile structure throughout their nucleus ranging from one to several tens of millimeter \citep{pouletetal2016}. Several numerical studies have been performed to explain these properties by postulating the formation of comets through the collapse of a pebble cloud \citep{Blumetal2017}.

Pebbles dissipate kinetic energy through collisions and gas drag, which leads to cloud contraction, as shown in pioneering work by \citet{WJJohansen2014,Wahlbergetal2017}. Depending on the total cloud mass, the initial size distribution can be altered significantly during collapse due to collisions that alter pebble mass, as well as varying terminal velocities for different pebble Stokes numbers. Even bouncing collisions can affect the initial imposed size distribution through pebble compaction, which aids the SI if the compaction timescale is faster than the radial drift timescale toward the star \citep{Loreketal2016,Loreketal2018}.

We take a more direct approach by solving the dynamical equation of motion of individual pebbles (swarms) that are subject to gas drag, mutual collisions, and gravity. In particular, we are interested in the collapse timescale and the final radial pebble size distribution over the collapsed core. We compare our findings with earlier work and observational data to increase understanding of the formation and final structure of comets and planetesimals.

The paper is structured as follows. In Section \ref{sec:msetup}, we present the general setup of our cloud collapse model. In Section \ref{sec:results}, we present the results of our simulations followed by the discussion in Section \ref{sec:discussion} and a summary and conclusions in Section \ref{sec:conclusion}.
\section{Model and setup}
\label{sec:msetup}
\subsection{Dynamical cloud evolution}
\label{sec:dynevol}
To perform the numerical models, we developed the code \texttt{implode}\footnote{The code is available at \url{https://github.com/astrojoanna/implode}. Version 1.0 of \texttt{implode}, which was used in this paper can be obtained from \url{https://doi.org/10.5281/zenodo.4395893}.}. We follow the collapse of a cloud of self-gravitating and colliding pebbles resulting from the SI \citep{YoudinGoodman2005}. We consider a spherically symmetric cloud of pebbles\footnote{We continue to refer to the particles in our cloud as pebbles since the definition of a pebble covers the range of particle sizes we consider in the simulations} with a distribution of masses, $f(m)$. The total cloud mass in pebbles $M_t$ corresponds to the mass of a solid planetesimal core of radius $R_c$ and density $\rho_{\bullet} = 1 \ \mathrm{g \ cm^{-3}}$. The pebbles are distributed evenly over the cloud's Hill sphere, which is defined as:
\begin{equation}
    R_\mathrm{H} = r_0 \left ( \frac{M_t}{3M_\star} \right )^\frac{1}{3}\quad,
\end{equation}
where $r_0$ is the orbital distance from the star with mass $M_\star$. The pebbles interact through mutual gravity, gas drag and mutual collisions. We extended our model to 3D to incorporate the effect of initial dispersion in the angular directions of the spherical cloud. To focus on the analysis of the influence of collisions and gas drag, we neglect shearing effects in our model. This will be incorporated in future studies to investigate the effect of increasing complexity.

The spherical symmetry imposed on our cloud allows us to model the gravitational interactions with the shell theorem: Pebbles only feel the gravitational pull of the mass distribution situated below them as if this mass were concentrated at the spherical center as a point source. Thus the gravitational acceleration of a pebble at radius $r_p$ is calculated from the mass situated at $r < r_p$ with respect to the cloud center of mass:
\begin{equation}
    \mathbf{f}_\mathrm{g} = -\frac{GM_\mathrm{enc} }{r_p^3}\mathbf{r}_p \quad,
\end{equation}
where $M_\mathrm{enc}$ denotes all mass that is closer to the center of mass than the corresponding pebble at position vector $\mathbf{r}_p$ in the cloud. We took the effect of gas into account with a simple drag prescription of the form:
\begin{equation}
    \mathbf{f}_\mathrm{d} = -\frac{1}{t_\mathrm{s}} \mathbf{v}_\mathrm{p}\quad,
    \label{eq:drag}
\end{equation} where $\mathbf{v}_\mathrm{p}$ is the velocity vector of the pebbles. For pebbles of radius $s$ and internal density $\rho_{\bullet}$ the stopping time is given by \citep{Whipple1972}:
\begin{equation}
    t_s = \left\{\begin{matrix} \displaystyle \frac{\rho_{\bullet} s}{\rho_{\mathrm{g}} v_{\mathrm{th}}} &                \textrm{Epstein regime: }  \quad s<\frac{9}{4}l_{\mathrm{mfp}} \\[5mm] 
    \displaystyle
    \frac{2 \rho_{\bullet}s^{2}}{9 \eta_d} & \textrm{Stokes regime:} \quad s \geq \frac{9}{4}l_{\mathrm{mfp}}\\ 
\end{matrix}\right.\quad,
\label{eq:tstop}
\end{equation}
\begin{figure}[t]
    \centering
    \includegraphics[width=.45\textwidth]{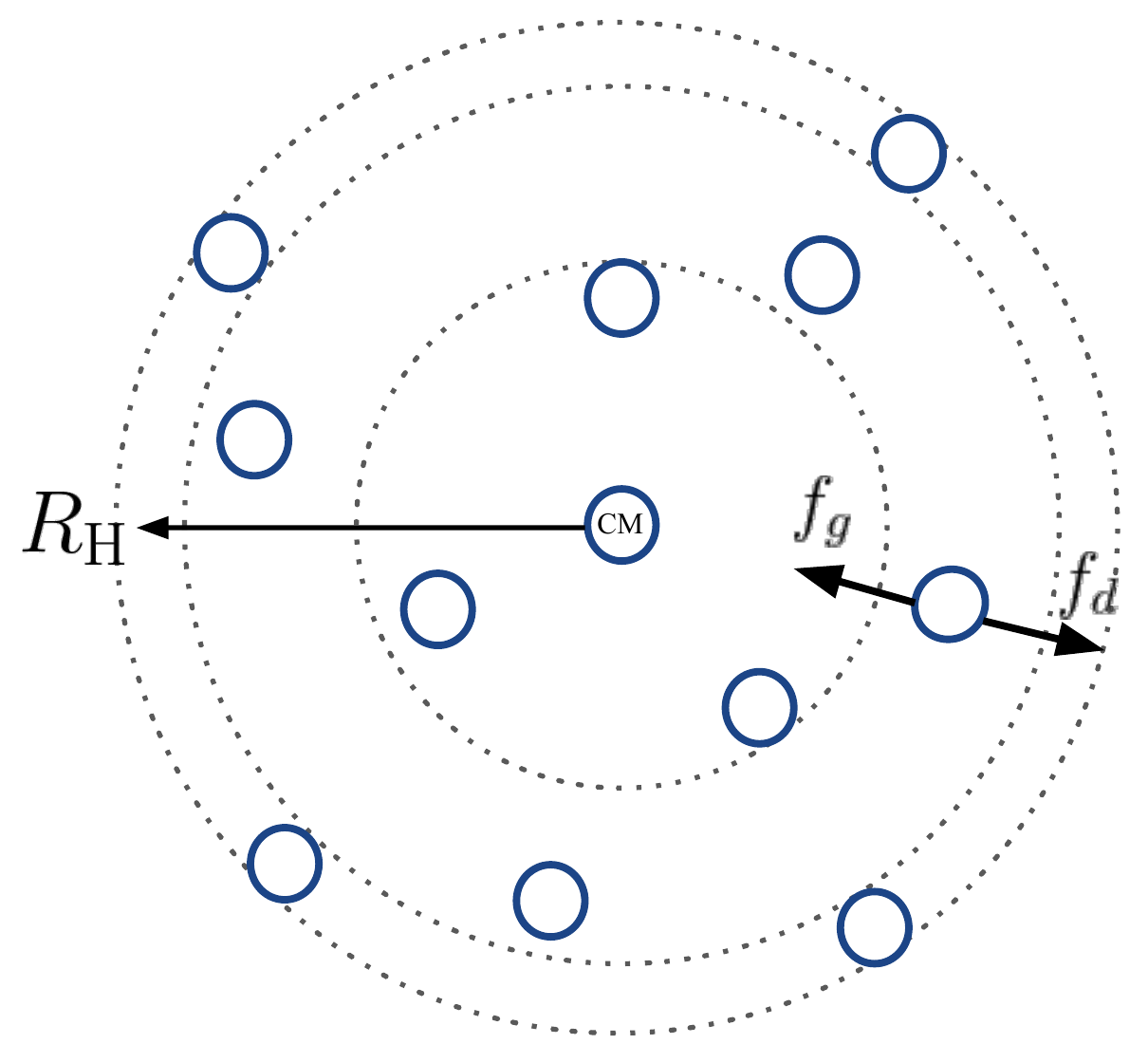}
    \caption{2D sketch of the spherical cloud collapse model with the center of mass (CM) in the origin. Pebble swarms (blue circles) are initiated over a Hill sphere with spherical velocities and positions. During evolution pebble swarms feel both gravity $f_g$ and gas drag $f_d$. The dashed shells indicate the zones for collision evolution. As the pebbles settle, zones are being rebuilt such that the number of swarms per zone never falls below a desired value $N_z$.}
    \label{fig:sketch}
\end{figure}
\begin{figure*}[t]
    \centering
    \includegraphics[width=1\textwidth]{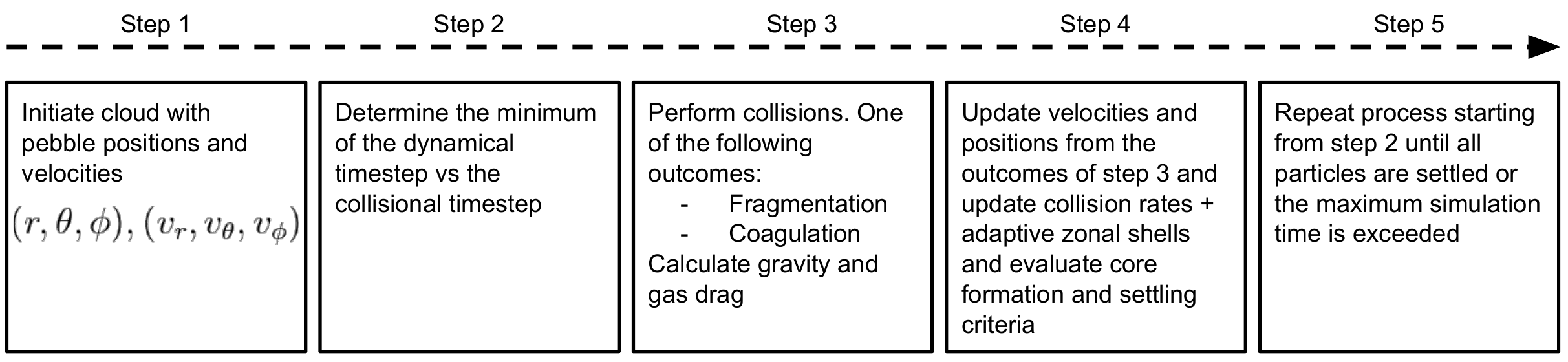}  
    \caption{Overview of the algorithm starting from cloud initialization at t=0.}
    \label{fig:chart}
\end{figure*}
where $v_\mathrm{th}$ is the thermal speed, $\rho_\mathrm{g}$ the mid-plane gas density, $l_{\mathrm{mfp}}$ the molecular mean free path, and $\eta_d$ the kinematic viscosity of the gas. The Epstein regime is the relevant regime for our range of pebble sizes since for 10 au the molecular mean free path is several meters. The Stokes number (St) relates the stopping time to one orbital timescale:
\begin{equation}
\mathrm{St} = t_s \Omega_0,
\label{eq:stok}
\end{equation}
with $\Omega_0$ the Kepler orbital frequency. The equation of motion governing the dynamics of these pebbles is given by:
\begin{equation}
    \frac{\mathrm{d}\mathbf{v}_\mathrm{p}}{\mathrm{d}t} = 
    -\frac{GM_\mathrm{enc}}{r_p^{3}}\mathbf{r}_p
     -\frac{\mathbf{v}_\mathrm{p}}{t_s}\quad.
\label{eq:eqofmotion}
\end{equation}

\subsection{The representative particle approach}
\label{sec:colls}
 The number of physical pebbles in the collapsing cloud is on the order of $10^{17}$ in our models. To be able to model them numerically, we model pebble collisions and advection using the representative particle approach \citep{Gillespie1975,ZsomDullemond2008} with much lower number of $N$ pebble swarms\footnote{We will refer to the representative particles in our simulation as pebble swarms or pebbles interchangeably.}, where $N \ll N_\mathrm{phys}$. Every pebble swarm represents $N_i$ physical pebbles with equal properties. The mass $M_\mathrm{swarm} = M_t / N$ of each swarm is equal and remains constant at all times. 
 
 We use an adaptive grid developed by \citet{Drazkowskaetal2013} to resolve the collisions locally in the cloud. We assume the cloud to be spherically symmetric. The natural way to subdivide local collision zones is through spherical shells over the cloud domain. The number of pebble swarms per spherical shell is kept above a desired value $N_z$ by rebuilding the shells if needed, to ensure enough resolution to perform collisions in every shell. In \Fg{sketch}, we provide a sketch of the cloud collapse model.
 \begin{table}[]
\begin{tabular}{lllll}
\hline
\hline
$r_0$ {[}au{]} & $R_c$ {[}km{]} & $R_\mathrm{H} / H_g$     & $M_t / M_\mathrm{Ceres}$ & $R_c / R_\mathrm{H}$   \\ \hline
               & 1              & $1.5 \ \times \ 10^{-5}$ & $2.65 \ \times \ 10^{-9}$ &                        \\
10             & 10             & $1.5 \ \times \ 10^{-4}$ & $2.65 \ \times \ 10^{-6}$ & $7 \ \times \ 10^{-4}$ \\
               & 100            & $1.5 \ \times \ 10^{-3}$ & $2.65 \ \times \ 10^{-3}$ &                        \\ \hline
               & 1              & $10^{-5}$                & $2.65 \ \times \ 10^{-9}$ &                        \\
39             & 10             & $10^{-4}$                & $2.65 \ \times \ 10^{-6}$ & $2 \ \times \ 10^{-4}$ \\
               & 100            & $10^{-3}$                & $2.65 \ \times \ 10^{-3}$ &                        \\ \hline
\end{tabular}
\centering
\caption{Overview of the simulation parameters: the disk radius $r_0$ and the planetesimal radius $R_c$, ratio of the radius of the initial pebble cloud (Hill radius, $R_{\mathrm{H}}$) and the disk scaleheight $H_g$, the cloud mass $M_t$ in the units of Ceres mass, and the ratio of the planetesimal radius to the Hill radius.}
\label{Tab:comparison}
\end{table}
 Collisions are then modeled with a Monte Carlo algorithm based on the total "particle in a box" collision rate in a zone
\begin{equation}
    R = \sum_{i}^{}\sum_{k}^{} R_{ik},
\end{equation}
where $R_{ik} = n_k \sigma_{ik} \Delta v_{ik}$ is the collision rate between a representative pebble from swarm $i$ and a nonrepresentative pebble of swarm $k$.

The collision time step is then chosen from the collision rate by drawing a random number $U \in [0,1)$ as:
\begin{equation}
    \label{eq:colltime}
    \delta t_c = -\frac{1}{R} \ln(U)\quad.
\end{equation}
After performing the collisions over the time step, the matrix of collision rates is updated accordingly. Since it is unlikely that representative pebble $i$ collides with any pebble from swarm $k$ we follow, we only update the rate of representative pebble $i$. In the case of mass change of pebble $i$ during a collision the physical pebble number represented by swarm $i$ changes to $N_i = M_\mathrm{swarm} / m_p$ where $m_p$ is the pebble mass and lower in case of fragmentation, and higher in case of coagulation. 

For a fragmentation threshold velocity below 10 m/s, or if bouncing collisions stop growth a very low Stokes numbers, the SI will not be triggered \citep{DrazkowskaDullemond2014,drazkowskaetal2016}. To meet the context of SI simulations, we limit ourselves in the possible collisional outcome, as a function of collisions velocity to sticking: $\Delta v \leq 10 \ \mathrm{m/s}$ and fragmenting:  $\Delta v > 10 \ \mathrm{m/s}$ \citep{Braueretal2008}. After a sticking collision the mass $m_i$ of the representative particle is updated to $m'_i = m_i + m_k$ and we use conservation of momentum to calculate its resulting velocity. In case of fragmentation, we assume that the mass of the original representative particle is distributed according to the power-law $n(m) \propto m^{-11/6}$, consistent with the MRN size distribution \citep{mathisetal1977}. We choose the new mass of the representative particle randomly from these fragments.
\subsection{Initial conditions}
\label{sec:incond}
We initiate a cloud of pebbles at a orbital distances of 10 and 39 au around a solar mass star, respectively. The pebbles are given positions and velocities (3D) $(r_p,\phi_p,\theta_p,v_{r},v_{\phi},v_{\theta})$ where we have adopted a spherical coordinate system. In the radial direction $r_p$, pebbles are placed such that the volume density is constant over the cloud. In the angular directions $\phi_p,\theta_p$, the positions are uniformly randomized over the angle domains $\phi \in [0,2\pi]$ and $\theta \in [\arccos{(-1)},\arccos{(1)}]$, respectively.

The SI in general relies on pebbles between minimum and maximum Stokes number $\mathrm{St}_\mathrm{min} \sim 2 \ \times \ 10^{-3},\mathrm{St}_\mathrm{max} \sim 10^{-1}$ \citep{Baietal2010, Yangetal2017}. We use an initial MRN size distribution between these Stokes numbers \citep{mathisetal1977}. For the chosen size distribution most of the cloud mass is found in the largest pebble sizes, while the lower sizes contain the most pebbles in number. Pebbles in these ranges of Stokes numbers are influenced by the gas on a timescale that is short compared to the cloud collapse timescale. We therefore initiate pebbles with their corresponding terminal velocity in the radial direction (see \api{termcoll} for details). For the angular directions $(v_\theta,v_\phi)$, following \citet{WJJohansen2014}, we initiate pebbles with a Maxwellian velocity dispersion:
\begin{equation}
dP(\Delta v) = \frac{1}{2 \sqrt{\pi}}\frac{\Delta v^2}{\sigma^3} e^{-\Delta v^2/4\sigma^2} d(\Delta v),
\label{eq:disp}
\end{equation}
with the dispersion $\sigma = \sqrt{2K_0/M_t}$. The random velocities are extracted from the initial potential $U_0$ and kinetic energy $K_0$ of the cloud, assuming initially virial equilibrium. A schematic overview of the cloud evolution is given in Figure \ref{fig:chart}. 
\begin{table*}[h!]
\centering
\begin{tabular}{l|l|l}
\hline \hline  
Parameter                                     & Description                               & Values                                \\ \hline
$\mathrm{St_{min}}$                           & Minimum Stokes number                     & $2 \times 10^{-3}$                   \\
$\mathrm{St_{max}}$                           & Maximum Stokes number                     & $10^{-1}$                            \\
$r_0$ {[}au{]}\                              & Orbital distance from star                & 10, \textbf{39}                               \\
$R_c$ {[}km{]}\                              & Core radius cloud                         & \textbf{1}, 10, 100, 500$^a$               \\
$\rho_{\bullet}$ {[}$\mathrm{kg \ m^{-3}}${]} & Internal density core and pebbles         & 1000                                 \\
$N$                                           & Number of representative pebbles          & $10^{4}$\\
tol                                           & Error tolerance dynamical evolution       & $10^{-7}$                            \\
$N_z$                                         & Minimum number of representative pebbles per zone & 200  \\
\hline
                             
\end{tabular}
\centering
\caption{Simulation parameters for the main results. All runs are performed with $N = 10^{4}$ representative pebbles except stated otherwise. Cloud parameters for the fiducial model are bold faced additionally. $^a$ 500 km core is only considered at 39~au distance.}
\label{Tab:tablepars}
\end{table*}
The gas in the disk is modeled according to the minimum mass solar nebula \citep{Weidenschilling1977B,HayashiEtal1985} with power law expressions for the gas temperature and surface density, respectively:
\begin{equation}
T(r_0) = 170\ \mathrm{K}\left ( \frac{r_0}{1 \ \mathrm{au}} \right )^{-1/2},
\label{eq:Temp}
\end{equation}
\begin{equation}
\Sigma(r_0) =1700 \ \mathrm{g \ cm^{-2}}\left ( \frac{r_0}{1 \ \mathrm{au}} \right )^{-3/2}\quad.
\label{eq:surfdens}
\end{equation} 

Simulations of the SI show that the planetesimals formed have a typical radius on the order of 50-100 km \citep{JohansenEtal2009,Schaeferetal2017}.
Kilometer-sized comets are unlikely to form via the SI in turbulent environments as clumps are easily broken up again. An increase in resolution of the SI simulations shows, however, that the smallest clumps decrease in size \citep{Johansenetal2011,SimonEtal2016}. Another reason that km sized object could survive is the fragmentation of a single clump into binary or satellite component \citep{Nesvorny2010, Nesvornyetal2019}. We consider three different cloud masses spanning the lower uncertain constraint on the core radius $R_c = 1$ km and the more certain upper constraints $R_c = 10$ km and $R_c = 100$ km. We locate the clouds at orbital distance 39 au (Kuiper belt distance) and 10 au (Saturn distance). We show the Hill radii compared to gas scaleheight $H_g$ and the core masses $M_\mathrm{Ceres}$ in Table \ref{Tab:comparison}

\subsection{Relative collision speeds}
\label{sec:relvels}
Modeling collisions accurately in this system is nontrivial because the velocities in the radial direction are systematic.  As the cloud collapses under the forces of gravity, pebbles are accelerated toward the center, with velocities moderated by the gas drag.  Whether pebbles are moving with terminal (due to friction) velocities or in free fall, these velocities strongly depend on location.

In a representative particle approach, we bin the radial structure of the cloud and consider collisions between pebbles in a radial bin.  We consider collisions between randomly selected pebbles in this bin, looking at relative velocities that may cause collisions.  Due to the systematic nature of the radial motions, and in particular the radial gradient in velocities, the relative velocity of any two pebbles will be overestimated unless the pebbles are located at exactly the same distance from the center of mass.  Because of this effect, we can expect that the results would significantly depend on the resolution in the model. To avoid this, we assume that the radial velocity of each pebble, for the properties of the collision, is the one the pebble would have when located exactly in the middle of each cell.  In addition to that, we still consider random velocities in the $\theta$ and $\phi$ direction.  These velocities are initially set from the initial conditions, are damped by the interaction with the gas, and are fed by the outcome of collisions. In \api{convergence} we show that this approach leads to robust convergence of the results as a function of resolution.


\begin{figure*}[t]
    \centering
    \includegraphics[width=.9\textwidth]{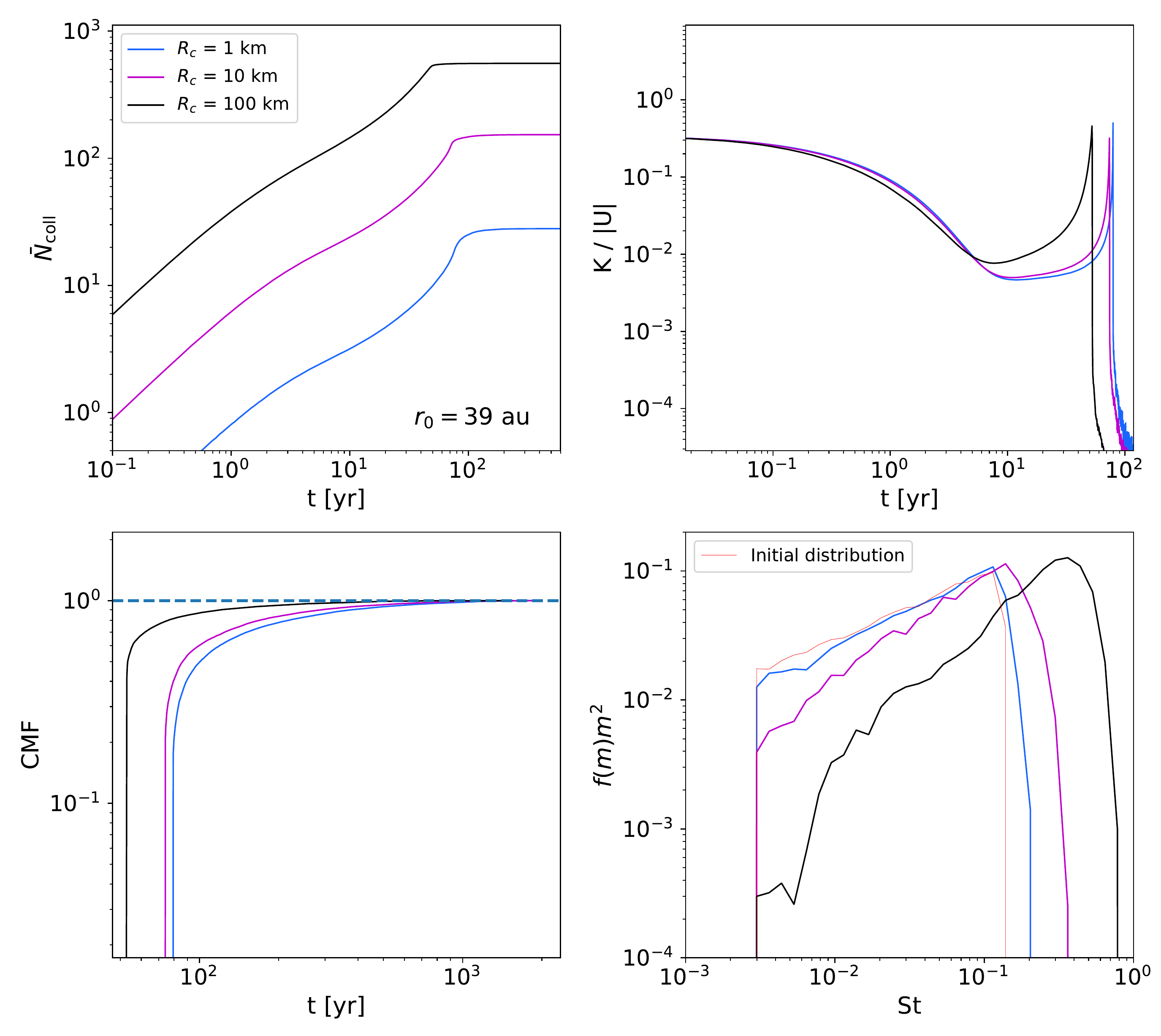}
    \caption{Simulation results for three different cloud masses corresponding to core radius $R_c = [1, 10, 100]$ km at 39 au from the central star. Top left panel: Average collisions per pebble $\bar{N}$, Top right panel: ratio of
kinetic to potential energy E = K/|U|, bottom left panel: CMF in time. bottom right panel: Mass weighted initial and final pebble Stokes number
distribution. The initial St distribution (red curve) is the same for all three cloud masses.}
    \label{fig:39au}
\end{figure*}

\subsection{Accretion conditions}
\label{sec:coreform}
Another numerical challenge of this system is that the core forming in the center is orders of magnitudes smaller than the starting size of the cloud. It is important to test if pebbles approaching the center of mass will get stuck there due to collisions right away, or if they will pass through the center and start a damped oscillation before finally settling toward the core.  We test this by checking at what point the center of the collapsing cloud becomes so dense that no pebbles will be able to pass through without suffering collisions. Basically, this is the same as asking, when the core will become optically thick. For our computation, we will assume that if the first pebbles arriving in the core become optically thick, a core has formed. After that, any pebbles arriving within a set distance of the center of mass will settle onto the core in the sequence of arrival. The rapid formation of a massive core was also observed by \citet{Wahlbergetal2017}. 

In all models, we are tracking the optical depth of one percent of the total cloud mass $M_t$ to ensure that this assumption is reasonable. In \api{convergence}, we demonstrate that this condition is matched for our fiducial model. If the innermost one percent of optically thick pebble swarms has settled, the cloud radius at this distance is adjusted to be the accretion radius for the outer swarms as $r_\mathrm{acc} = r_{\tau > 1}$. Here $r_{\tau > 1}$ is the position where the first one percent inner mass reached $\tau_\mathrm{in} > 1$. Pebbles are therefore accreted if they meet the condition:
\begin{equation}
r_p - r_\mathrm{acc} < 0.
\label{eq:acccond}
\end{equation}
with which we settle pebbles in the core with uniform density, leading to growth in core radius according to $r_c = (3N_\mathrm{sett} m_p / 4\pi\rho_{\bullet} )^{1/3}$, with $N_\mathrm{sett}$ the amount of pebbles settled. We also keep track of the sequence with which pebbles settle in time to preserve information about the final size distribution of pebbles over the radial extend of the core layers.

\subsection{Numerical scheme}
We integrate the equation of motion of the pebble trajectories while we simultaneously resolve the mutual pebble collisions using the representative particle approach. The equation of motion is integrated in 3D with a Runge-Kutta Fehlberg variable step scheme \citep{Fehlberg1969} with an error tolerance $\mathrm{tol}= 10^{-7}$. The time step following from the RKF45 solver, $\delta t_\mathrm{EOM}$ is compared with the longest time step we can afford to include collisional evolution reasonably well, $\delta t_c$, which is calculated as a minimum of timesteps reported by each radial zone: the average timestep between two consecutive collisions multiplied by the number of particles in the zone. We pick the minimum time step from this to ensure we resolve both the dynamical and collision part of the cloud evolution:
\begin{equation}
\Delta t = \min [\delta t_\mathrm{EOM}, \delta t_c]\quad.
\end{equation}

If pebbles are within the accretion radius $r_\mathrm{acc}$ we place them in the core accordingly. The $N_\mathrm{sett}$ pebbles are considered as settled do not participate in collisions and in the further cloud evolution anymore.

\section{Results}
\label{sec:results}
In \Fg{39au} and \Fg{10au}, we present results for the cloud located at 39 au (with $R_c = 1$ km the fiducial value) and 10 au orbital distance from the central star, respectively. An overview of the parameter study is given in Table \ref{Tab:tablepars}. The total number of representative pebbles is $N = 10^{4}$. The core density and internal pebble density are $1\ \mathrm{g \ cm^{-3}}$ and the initial size distribution for all three cases corresponds to $[\mathrm{St_{min}} = 2 \times 10^{-3},\mathrm{St_{max}} = 10^{-1}]$.

\begin{figure*}[t]
    \centering
    \includegraphics[width=.9\textwidth]{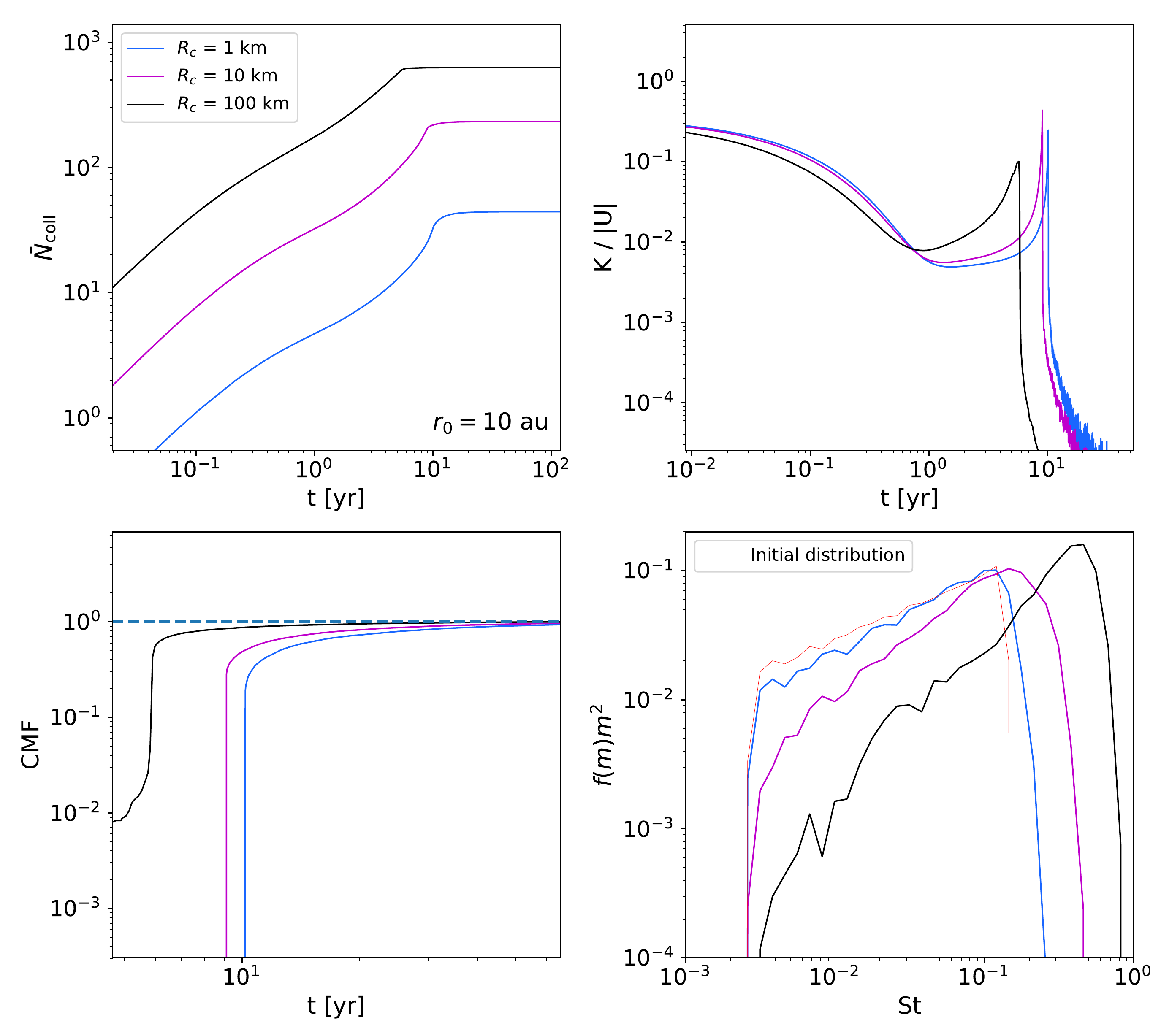}
    \caption{Same results as shown in \Fg{39au} but now for an orbital distance of 10 au for $N = 10^{4}$ in all three cases. The most notable changes are a significantly shorter collapse time, increasing number of collisions due to smaller Hill radii, and a gradually increasing CMF. Growth can be significant as a result of sticking collisions for 100 km core radius. }
    \label{fig:10au}
\end{figure*}
Starting with the 39 au case, we show the average number of collisions per representative pebble $\bar{N} = N_\mathrm{coll} / N$, with $N_\mathrm{coll}$ in \Fg{39au}, top left panel.  For increasing core radius $R_c$, $\bar{N}$ increases due to the higher number density in smaller pebbles. An estimate of the increase rate of $\bar{N}$ for larger core radius is given in \api{avcolls}. Starting from $t = 0$, the average collisions number $\bar{N}$ increases and approaches a constant value after some time. This is explained as follows. The most extreme difference in radial velocities is found between the largest and the smallest pebbles. The local collision rates in the radial direction are fed predominantly by this extreme case. We typically observe that the most massive pebbles form the first core, leading to a large drop in $\Delta v_r$ for the smaller active pebbles. By the time when $\bar{N}$ is constant, the core has formed from the inner one percent of cloud mass with $\tau_\mathrm{in} \sim 1$. This sets the accretion radius for the active remaining pebbles in the cloud. Collisions that would normally occur in the over-dense region within the accretion radius are now excluded. While this could lead to some collisional processing in the collapsing shells, the high optical depth guarantees that pebbles cannot change sequence within $r_\mathrm{acc}$.

The top right panel of \Fg{39au} shows $E = K/|U|$, the ratio of the total kinetic energy to the total absolute value of the potential energy of the pebbles in the cloud. We observe that the system never reaches anything close to the virial equilibrium, which would correspond to $E = 0.5$, except for the rapid interval where freefall restricts addition of further energy at the peak of E (top right panel of \Fg{39au}, global peak). Instead, the formation of a massive core happens before pebbles can cross the central region of the cloud. This prevents pebbles from oscillating through the center of the cloud and causes the kinetic energy of the system to drop to zero if all pebbles have settled.

The evolution of $E$ is explained as follows. Starting at $t=0$, the kinetic energy of the individual pebbles is fed by the virial assumption for initial dispersion given in \eq{disp}. As time evolves, the most massive pebbles typically form the first core characterized by the sharp global maximum in $E$. The fall times of the massive pebbles are approximately equal in the narrow bigger size range (\fg{falltime}) and they are only weakly coupled to the gas. The sharp global maximum in $E$ is a combination of the potential energy efficiently being converted to kinetic energy due to their high terminal velocities and a collective, near instantaneous, arrival at the center of mass. Shortly after the maximum, these massive pebbles have formed the core and no longer contribute to the $E$ evolution, explaining the steep drop in $E$. The difference in fall times for smaller pebbles increases rapidly with decreasing size since $\frac{\mathrm{d} t_f}{\mathrm{d} \mathrm{St}} \propto -1/\mathrm{St}^2$. This causes a much more gradual decrease in $E$. If pebbles have reached the accretion radius, they are frozen in the core. The total energy therefore declines with respect to the energy at t = 0 since we exclude them from the further energy evolution of active pebbles. This explains the fluctuation in total energy during the decline to E=0, every time a pebble is settled. In \api{convergence}, we present a resolution study for different representative pebbles. The most telling parameter is the average number of collisions experienced by a pebble, and we can see that, for sufficient resolution, $\bar{N}$ remains the same for increasing $N$. 
\begin{figure*}[t]
    \centering
    \includegraphics[height=7.5cm]{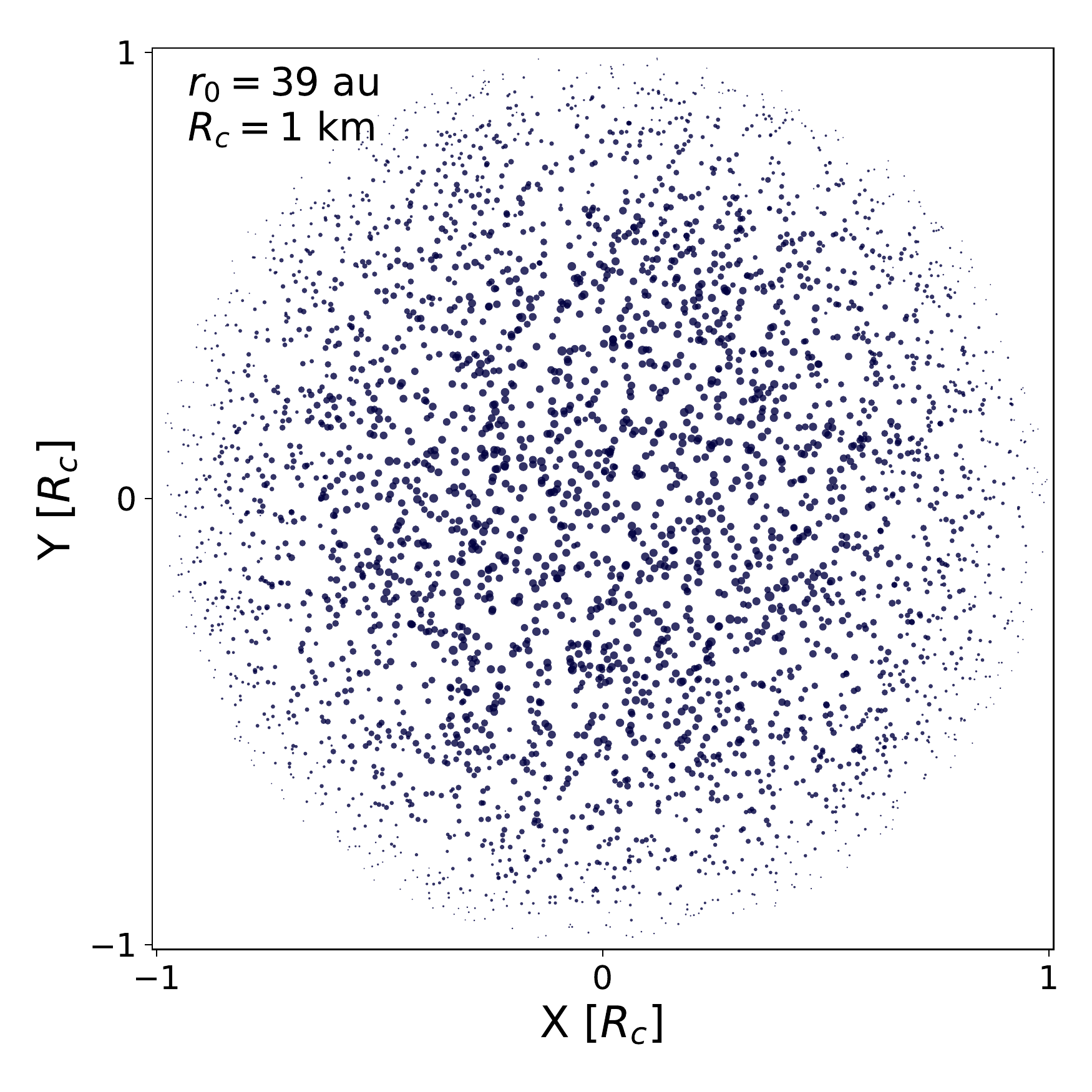}
    \includegraphics[height=7.5cm]{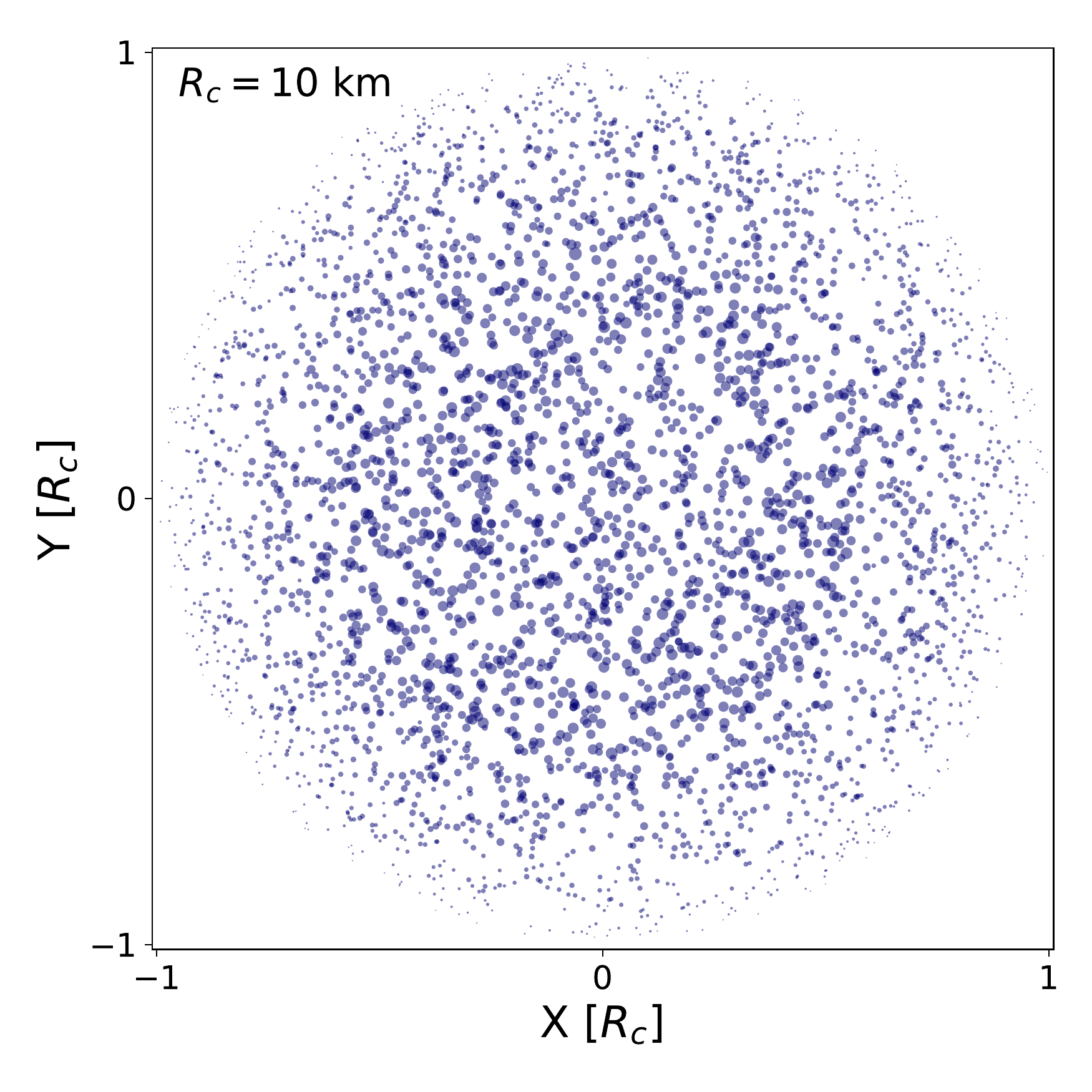} \\ 
    \includegraphics[height=7.5cm]{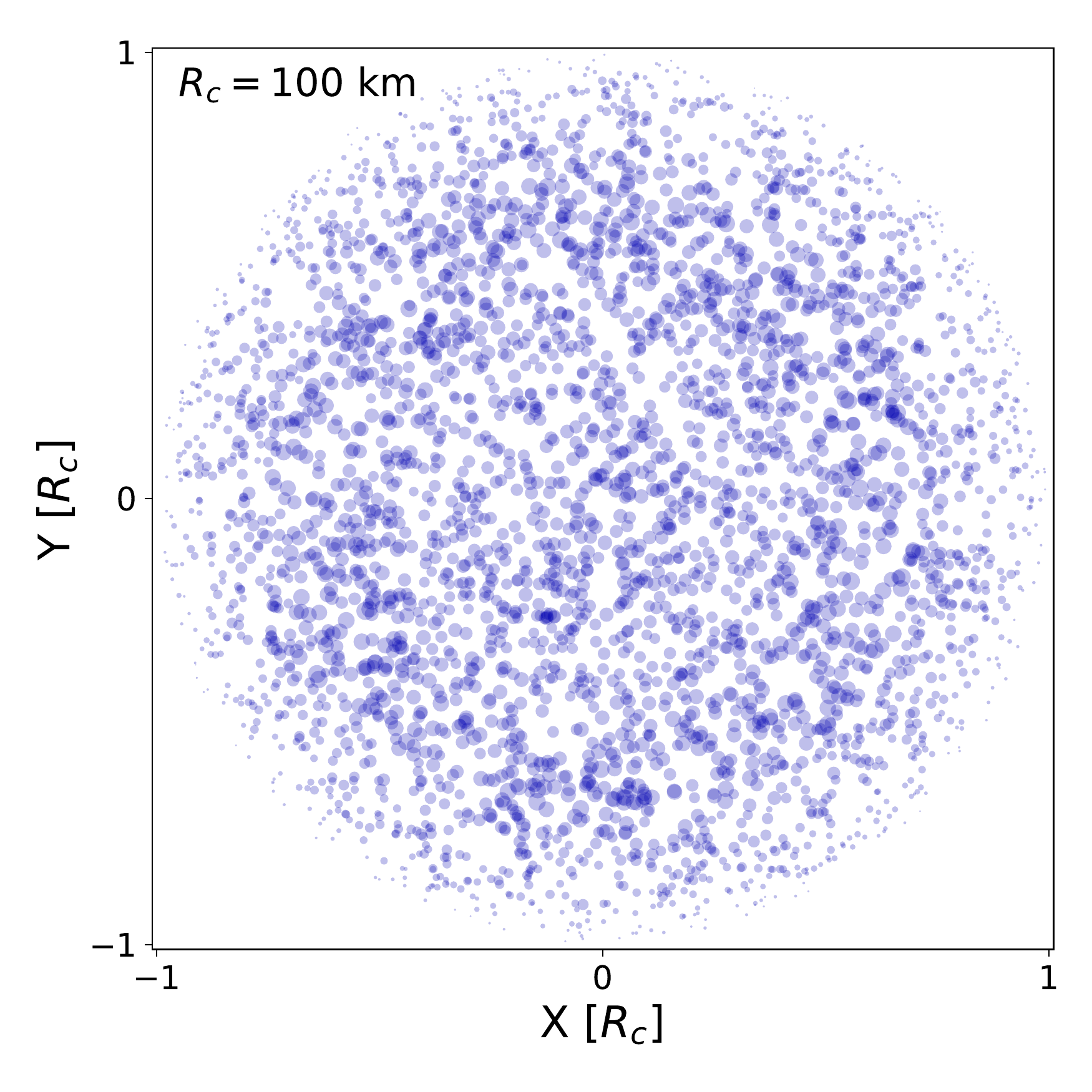}
    \includegraphics[height=7.5cm]{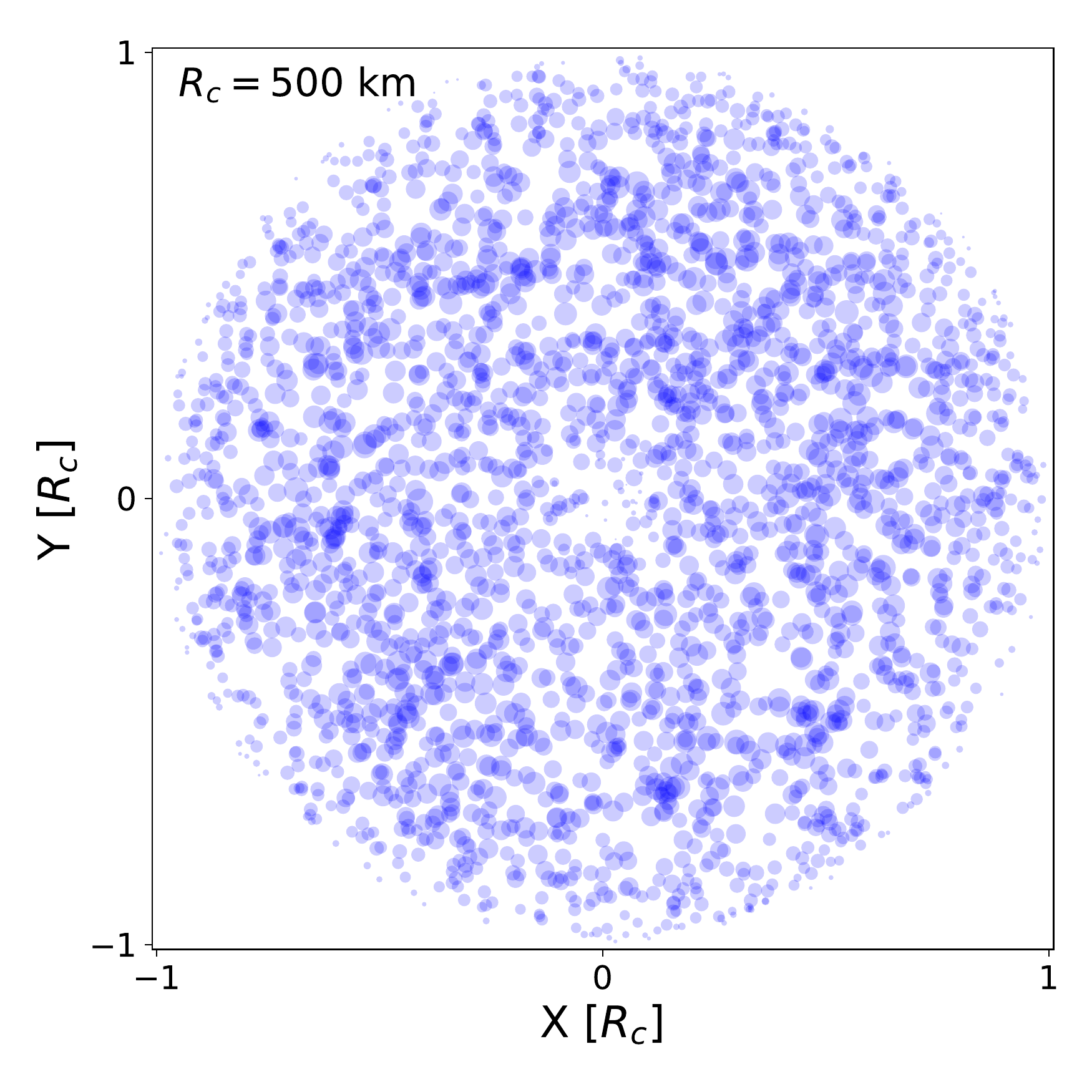}
    \caption{2D slices in the XY plane of the final core formed from the collapse on 39 au normalized in units of $R_c$. The z-dimension has been flattened to one percent of the core radius. Pebbles are indicated with the circles and scale from smallest circle to largest circle with $\mathrm{St_{min}},\mathrm{St_{max}}$ resp. Top left panel: Core structure for $R_c=$ 1 km. The inner core is composed of primarily $\mathrm{St_{max}}$ pebbles gradually decreasing toward $\mathrm{St_{min}}$ toward the core surface. Top right panel: Ditto for 10 km. Bottom left panel: Increasingly more smaller pebbles mix in between the massive inner core for 100 km core radius. Bottom right panel: For $R_c = $ 500 km there is a more diverse mixture of pebble sizes in the core. }
    \label{fig:corestruct39au}
\end{figure*}
The core mass fraction (CMF; \Fg{39au}, bottom left panel) indicates the fraction of mass in pebbles that has settled with respect to the total cloud mass (bottom-left panel). The steep vertical increase in the CMF can be understood in the same way as the global peak in $E$. The massive pebbles all fall to form the core together, leading to a sudden increase in the CMF. The more gradual infall of the smaller pebbles leads to a smoother and more gradual increase in the last few percent of the CMF referring to the same explanation as for the behaviour in $E$ after the global peak. For massive clouds that are optically thick in the inner region already at $t=0$, the CMF may be nonzero already from the start since we form the core at $t=0$ in this case. This is only observed for the 10 au case for $R_c = 100$ km (\Fg{10au}, bottom left panel).
\begin{figure*}
    \centering
    \includegraphics[height=7.2cm]{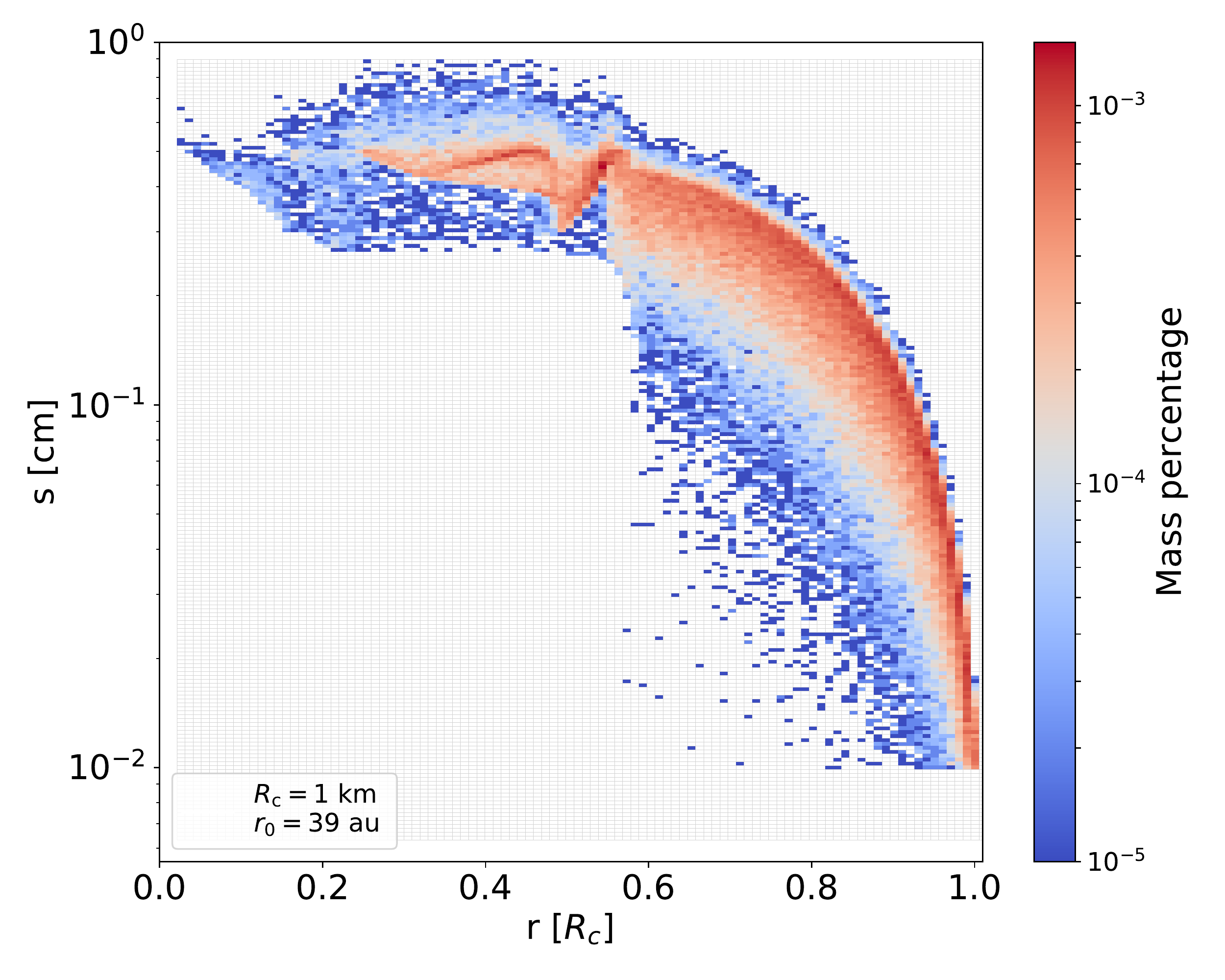}
    \includegraphics[height=7.2cm]{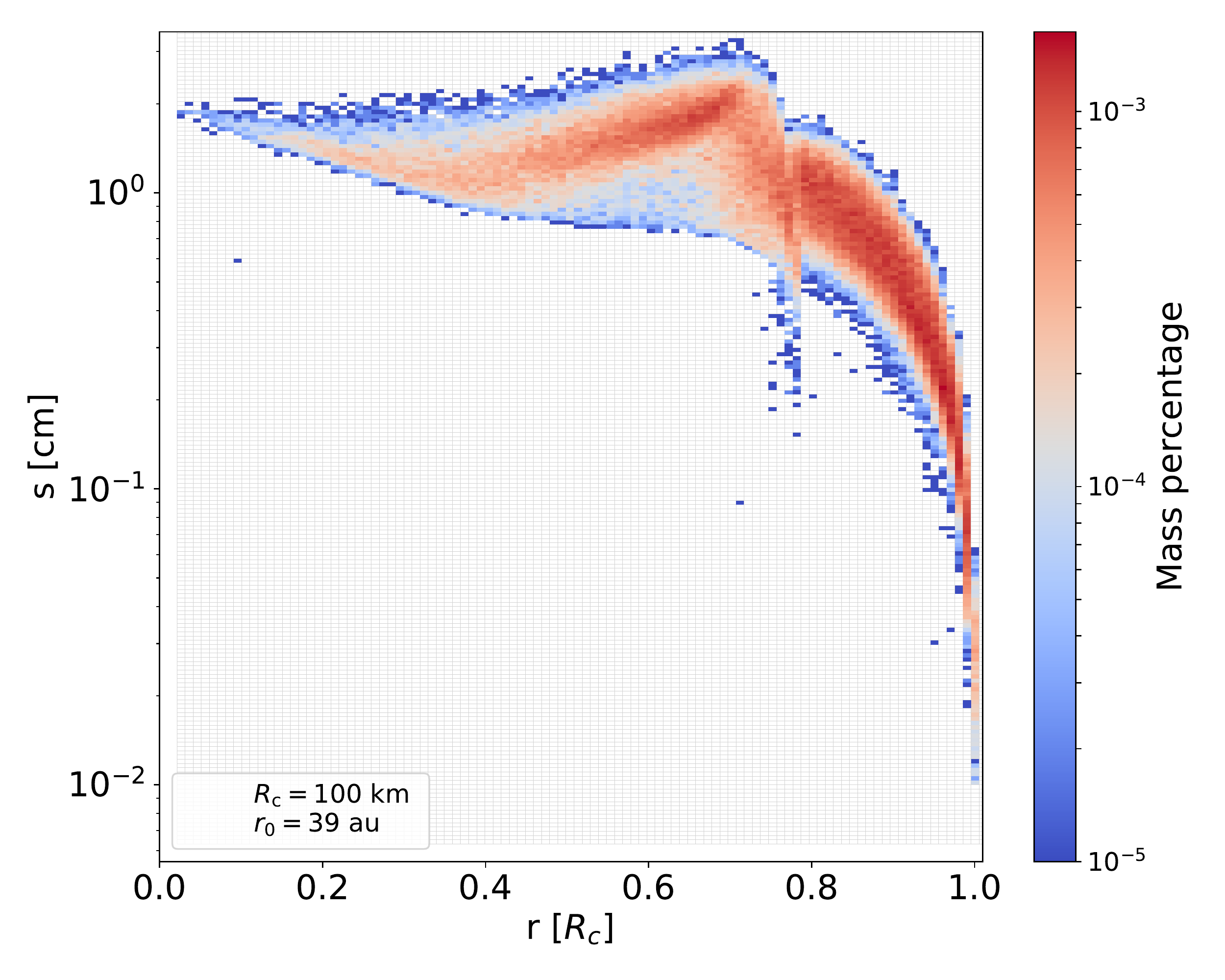} \\ 
    \includegraphics[height=7.2cm]{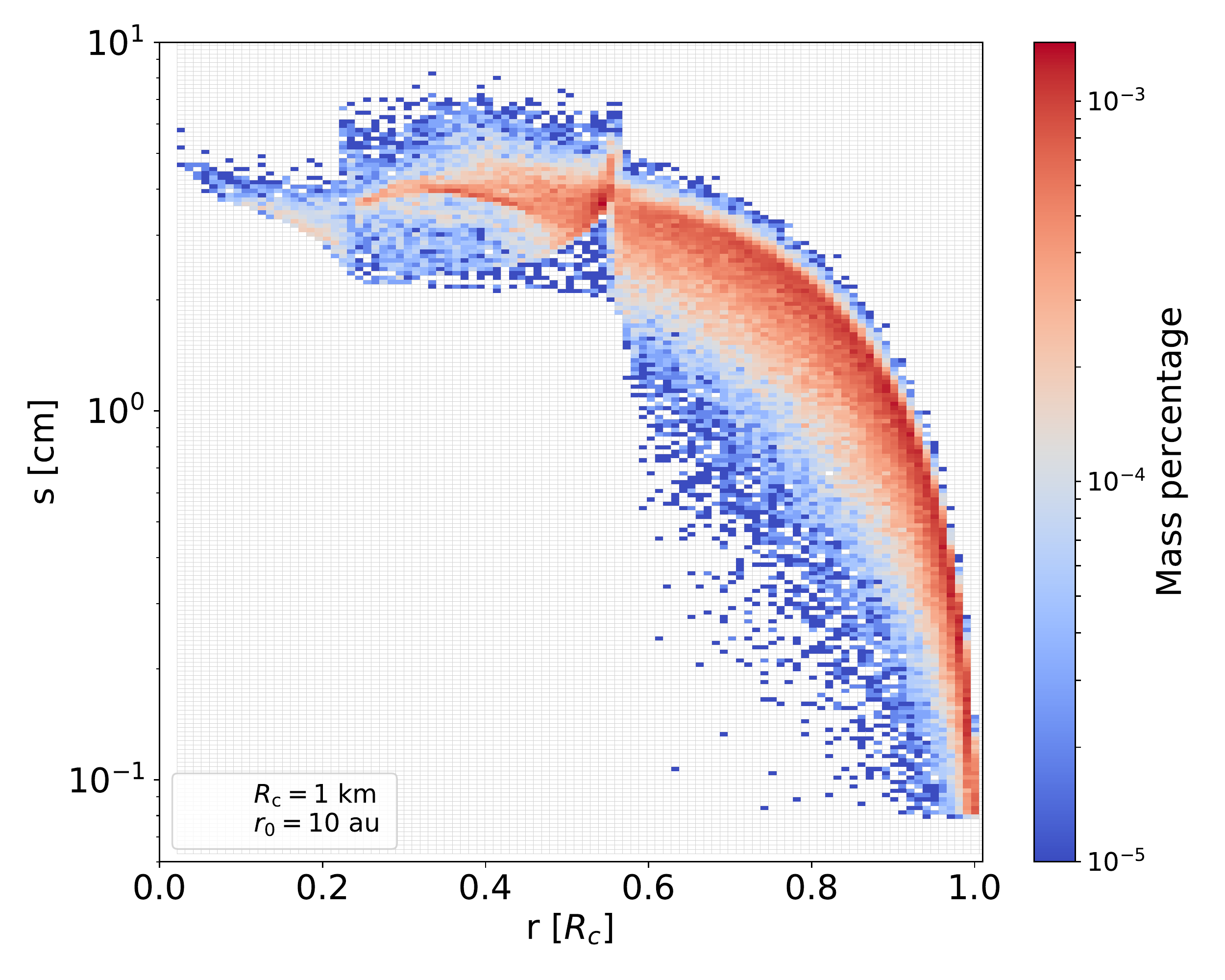}
    \includegraphics[height=7.2cm]{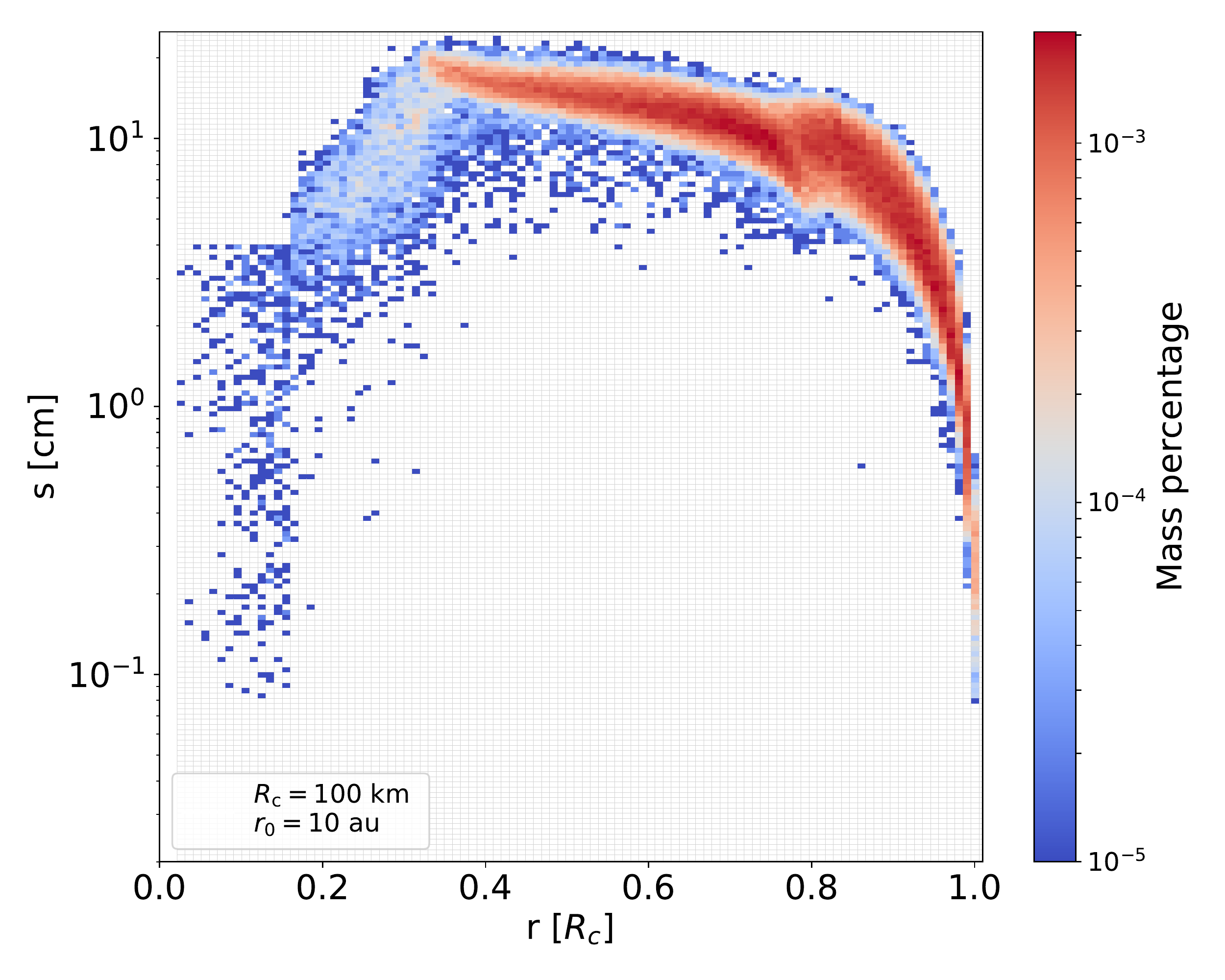}
    \caption{Contributions to the total core mass for 1 km and 100 km at 39 au (upper left and right respectively) and 1 km and 100 km at 10 au (bottom left and right respectively) for a high resolution run of $N = 10^{5}$ pebble swarms. The y-axis indicates the size range of pebbles and the x-axis the distance from the core center. All the pebbles in a certain pixel with width $\Delta r$ and height $\Delta s$ are counted and weighted with respect to $M_t$, the total core mass. The mass contribution is shown by the color bar.}
    \label{fig:heatmaps}\end{figure*}
The initial and final size distribution is shown in \Fg{39au}, the bottom right panel. For low mass clouds ($R_c = 1$ km) the average collisions per pebble $\bar{N} \sim 10$ shows that pebbles are hardly affected by collisions, even more so since collisions are dominated by larger pebbles colliding with smaller ones. This reflects back in the final size distribution being almost identical to the initial one. As cloud mass increases, collisions become more important and the final size distribution is shifted increasingly toward higher Stokes numbers. This shows that the massive pebbles grow larger by sweeping up the smaller ones through coagulation since the small pebbles from the initial distribution are depleted in the final one. It is worth noting that we are not in the classical situation of coagulation happening over orders of magnitude in size with a narrow size distribution, in which even a low number of collisions can cause significant growth. Rather, we have large pebbles collecting some smaller ones. So even in the case of collecting pebbles of the same size as the original one, ${\bar N}$ collisions only increase the mass by at most a factor of ${\bar N}$. $\bar{N} \sim 10$ means a maximum change of size, and the proportional Stokes number, by a factor of $10^{1/3}\approx 2$. This is consistent with the Stokes number distribution presented in the bottom right panel of \Fg{39au}.

For the computations at 10 au, we show the same parameters as the 39 au case in \Fg{10au}. The most noteworthy change in results is a shorter fall timescale for the cloud due to decrease in cloud Hill radius by a factor of four. The same mass is now initially distributed over a smaller region. Average collisions per pebble inevitably increase with respect to the 39 au case (\Fg{10au} top left panel). This leads to a more rapid increase of the optical depth in the central region. For the 100 km case, the inner active core mass already becomes optically thick at t = 0. The CMF increase is therefore, after an initial jump, more gradual already at t = 0 since the full mixture of the initial size distribution contributes immediately by falling through the  accretion radius $r_\mathrm{acc}$ (\Fg{10au}, bottom left panel). An additional overview of the main parameters used in the simulations is given in Table \ref{Tab:tablepars}.
\subsection{Fragmentation events}
We rarely observe fragmentation happening in the simulations, even for very massive clouds in which pebbles should, in principle, reach terminal velocities exceeding the fragmentation threshold $\Delta v = 10$~m/s. To find out why this is the case, we look at the maximum terminal velocities that pebbles can reach before arriving at the accretion radius. We estimate the accretion radius where the optical depth of the first inner one percent of the cloud mass $M_t$ exceeds unity (see \api{convergence} for details).
\begin{equation}
    \tau_\mathrm{in} = \frac{N \sigma}{4 \pi r^2}\quad, 
    \label{eq:opdep}
\end{equation} 
with $N = 0.01M_t / m_p$ the physical number density of pebbles contained in the inner one percent of total cloud mass, $\sigma = \pi s^2$ their physical cross-section and $4\pi r^2$ the surface of the sphere enclosing the $0.01M_t$ in pebbles. If $\tau_\mathrm{in} > 1$, the contracting core is optically thick and becomes impenetrable for outer mass. Solving for the accretion radius $r_\mathrm{acc}$ in Equation \ref{eq:opdep} gives
\begin{equation}
    r_\mathrm{acc} = \left ( \frac{0.01 R_c}{4 s} \right )^{1/2}\quad.
    \label{eq:racc}
\end{equation}
Filling in this expression for the terminal velocity of a pebble at this particular radius $v_t = G 0.01M_t \mathrm{St} \Omega_0^{-1} / r_\mathrm{acc}^{2}$ yields a critical terminal velocity $v_{t,*}$ at the accretion radius of:
\begin{equation}
    v_{t,*} \sim 60 \ \mathrm{m \ s^{-1}} \ \mathrm{St}^2\quad.
    \label{eq:vcrit}
\end{equation}
This expression is independent of disk radius for the specific power law $n = -3/2$ of the gas surface density (\eq{surfdens}) and only depends on the square of the Stokes number. For other power law profiles of the gas surface density there will only be a weak dependence.

To summarize this result: Fragmenting collisions between pebbles are difficult to achieve due to optical depth of the cloud exceeding unity before the collapse has been finalized (\api{convergence}, \fg{test100km}, bottom right panel). The cloud becomes optically thick at cloud radii where the terminal velocity needed for fragmentation cannot be reached. We expect (even though we  are unable to track these collisions for numerical reasons) that pebble collision rates beyond this region prevent pebbles from reaching high speeds due to rapid damping. We propose from these findings that fragmentation events in collapsing pebble clouds are rare and comets are composed from primordial building blocks. Indeed this result is consistent with models of the surface and interior of comet 67P that reveal an active surface layer of $\sim$ 1 meter and primordial pebbles below that \citep{Capriaetal2017}. This strengthens the claim that small solar system objects are in general primordial in nature \citep{Bottkeetal2005}.

\subsection{Final core structure}
In \Fg{corestruct39au}, we show 2D XY slices through the center of the final core for the fiducial model with $R_c =$ 1 km, $R_c =$ 10 km, $R_c =$ 100 km and an extreme case of $R_c =$ 500 km at 39~au. In \Fg{heatmaps}, we present the contribution of the radial core layers to the total core mass for 1 km and 100 km core radii, at 39 and 10 au. In general, the resulting core structure for low mass clouds ($R_c = 1, 10, 100$ km for 39 au and $R_c = 1, 10$ km for 10 au) can be explained as follows. The sequence of arrival of pebbles is dominated by their corresponding terminal velocities. Pebbles are found further from the center of mass of the core for decreasing terminal velocity. The resulting core structure is characterized by massive pebbles in the center, followed by increasingly smaller pebbles toward the core surface (\fg{corestruct39au}, top left, top and bottom right panel).

The massive pebble swarms form the core, and while they dominate the mass, they are fewer in number. This explains that the massive inner core region is only a small mass contribution with respect to the total core mass (\Fg{heatmaps}, the top and bottom left panels). If we look closer toward the core surface, we observe that contributions to the total mass become higher. The main reason for this is that the pebble number density increases for decreasing Stokes number. It then follows that more and more pebbles fall at approximately the same time to the core. The total amount of pebbles in one pixel of width $\Delta r$ and height $\Delta s$ increases toward the surface since the mass of one swarm is constant.

In general, in all cases in \Fg{heatmaps} there is a region with pebbles larger than the pebbles in the core while they have ended up further outwards. These are pebbles that grew through coagulation while the largest pebbles from the initial size distribution already formed the core. The sequence of increasingly larger pebbles with increasing core depth is not fully observed in the 100 km (10 au) and 500 km (39 au) core (\fg{corestruct39au} bottom panels, \fg{heatmaps} bottom right panel). For the largerst core mass $M_t$ at a given orbital distance, the inner region becomes optically thick at an earlier stage due to $\tau _\mathrm{in} \propto M_t$. This sets the accretion radius at an increasingly higher fraction of the Hill radius for larger $M_t$ ($r_\mathrm{acc} \sim 0.2 R_\mathrm{H}$ for $R_c = 500$ km).

The observed accretion sequence of pebbles in the massive cores reflects the initial placement of the pebbles. An extreme example is the abovementioned run with 500~km planetesimal at 39~au, for which the inner part of the pebble cloud is optically thick already at $t=0$. All the pebbles within the accretion radius $r_\mathrm{acc} \sim 0.2 R_\mathrm{H}$ are directly settled including the small pebbles. Additionally, since pebbles now accrete at $r_\mathrm{acc} \sim 0.2 R_\mathrm{H}$, there is less time to differentiate the fall timescales of the small and large pebbles resulting in a more randomized accretion sequence for the outer core too. This observation implies that a significant core might already form during the SI phase for massive clouds, before the gravitational collapse of the full cloud is even starting. 

An important aspect to mention is our choice of initial cloud radius. As mentioned by \citet{Wahlbergetal2017}, the choice of distributing the cloud mass over its corresponding Hill radius leads to self-similar collapse for given orbital distance $r_0$. The reason for this is that free-fall timescales as well as terminal-velocity fall timescales are independent of cloud density and mass. We therefore predict that comets and planetesimals are composed of a solid core formed by primordial massive pebbles. The outer layers are constructed of increasingly smaller pebbles toward the surface of the body. Especially for comets the outer layers would be stripped down easily over tens of eccentric orbits around the central star, revealing the inner core of primordial centimeter sized pebbles \citep{pajolaetal2017,arakawaohno2020}. 

\section{Discussion}
\label{sec:discussion}
One of the key assumptions that goes into out model is the initial setup of the cloud, just before collapse.  In reality, there will be an organic transition from a phase in which pebbles are being collected into a clump through the SI into a phase where gravity dominates and triggers the gravitational collapse that we studied in this paper. The initial setup concerns the spatial distribution of pebbles as well at the initial velocity distribution of those pebbles. The velocity dispersion of pebbles during the SI phase are unknown and not easy to obtain without currently unfeasible ultra-high resolution simulations of the SI. Processes such as gas coupling, turbulence, dust to gas feedback and random Brownian motion play an important role in determining the right dispersion profile. If turbulence is important in the cloud, small-scale clumps resulting from the SI are easily destroyed again \citep{Johansenetal2011,KlahrSchreiber2020} where the latter authors do note that this effect vanishes likely as gas depletes in the later disk stage. The typical speed induced by turbulence is given by $v_\mathrm{turb} = \sqrt{2\alpha \mathrm{St}}c_s$ for St $< 1$ \citep{Ormelcuzzi2007} with $\alpha$ the turbulence parameter. For a standard value of $\alpha = 10^{-3}$ and St = $10^{-2}$ we obtain $v_\mathrm{turb} \sim 0.5$ m/s. 
Since the terminal velocities of the smallest pebbles are of similar order of magnitude (see \eq{vcrit}), we do think turbulence may be an important effect to take into account in future follow-up studies. In general for the final collapse phase of the global SI (which we track), starting from a cloud on $R_\mathrm{H}$, we believe that the terminal settling velocities are dominant due to the strong gas coupling regime that we consider. We therefore use terminal velocities to initialize the radial velocities. For the angular dispersion we use the virial dispersion profile based on the initial kinetic and potential energy of the cloud.  It is not certain how realistic this assumption is, but since the gas drag rapidly diminishes the non-radial velocities, we believe that the influence on our results is small. Further investigations on velocity dispersion inside SI clumps would be important.

We use spherical symmetry with the shell approach for gravity. It is well known that small radial perturbations in the density/velocity profile of a collapsing gas cloud lead to an instability in which the density peaks at certain cloud radii \cite{Brennerwitelski1998}. This appears to be a consequence of forcing particles to collapse at different times for a point particle collapse, destroying the self-similarity of the free-fall solution (a simple proof is given in \api{shells}). Indeed, we observe the same phenomenon in runs with narrow or single sized initial pebble size distributions. For broad size distributions we do not encounter this instability due to the much higher difference in pebble fall times. It is still unclear to us whether this instability is an artifact or an intrinsic property of the self-similar cloud collapse solution.

One physical and important effect not included in our models is that of initial cloud rotation. The presence of rotation leads to rotational support of the cloud and might prevent pebbles from collapsing to the center of mass. However, if mutual pebble collision provide an effective viscosity, the pebbles could still collapse to a single core. Results in the shearing sheet approach also show that rotation is important in answering the question if a single object forms from the collapse or a binary system \citep{Nesvorny2010,Robinsonetal2020}. We plan to incorporate rotation and its implications in future simulations.

The high solids-to-gas ratio in the pebble cloud can lead to entrapment of gas during collapse. Modeling compressible gas is beyond the scope of this paper. We speculate that hydrostatic effects might slow down collapse. On the other hand, the gas might rapidly escape outward before the core density becomes critical. We recognize the importance of this effect and it should be investigated further in future follow-up.

Earlier work came to the conclusion that the core of kilometer sized bodies is formed  initially by mid-sized (centimeter) pebbles due to optimal energy dissipation by the gas \citep{Wahlbergetal2017}. Smaller pebbles take more time to settle, and larger pebbles do not lose enough energy through gas friction.
This result is indeed expected in a virial approach in which pebbles oscillate freely through the center of mass during contraction, so that collisions dominate energy losses. The oscillation is then dampened only slowly for pebbles that hardly feel the gas, and most quickly for the pebbles with high collision rates. In our model, the virial equilibrium is not observed in any of the simulations. Instead the core forms right away from the pebbles that are most massive, before other, smaller, pebbles reach the center of mass, as the cloud becomes optically thick. The largest pebbles suffer many collisions before crossing the center of mass for an oscillation. In this way not only pebbles with the ideal damping behavior through gas friction, but also larger pebbles are incorporated into the core right away. 
On the other hand the smallest grains will have low terminal velocity and reach the core last, leading to a planetesimal/comet surface dominated by small particles. The gas is clearly an important aerodynamic sorting mechanism for the final core structure, particularly for a unequal initial size distribution. The effect of gas was not considered important in some earlier work due to collision timescales being much lower than stopping timescales \citep{Nesvorny2010,WJJohansen2014}. 
However, in both these studies, the estimates of collision times was based on the assumption of randomly oriented, virial-like velocities. In our simulations, as the random velocity components get initially dampened by gas and maybe also collisions, the cloud enters into a more organized collapse where gas friction turns out to be the dominant factor.  Collisions still happen, but in a realistic situation where there will be some spread in Stokes numbers, these collisions will now be dominated by large particles sweeping up smaller ones in systematic motion. Therefore, we find that gas friction remains a key ingredient in the collapse model.


\section{Summary and conclusions}
\label{sec:conclusion}
We summarize the core collapse model that we have developed as follows. We track the evolution of a local pebble cloud that is graviationally bound resulting from the SI. The pebbles are initially placed over a sphere with radius $R_\mathrm{H}$ such that the mass density of the cloud is constant everywhere. The pebbles are subject to gas drag, mutual collisions and self-gravity. The initial radial velocities of the pebbles is set to their corresponding terminal velocity. This is a reasonable assumption since any random component in the radial direction is dampened to the terminal velocity due to the efficient gas coupling of the pebbles. In the angular directions we randomize the velocities using a virial dispersion profile. Pebbles are picked from an initial MRN size distribution between $\mathrm{St_{min}}, \mathrm{St_{max}}$, in accordance with the SI. 

Collisions are fragmenting or sticking, depending on the relative pebble velocities. Collisions are resolved using a Monte Carlo algorithm in which we statistically pool pebbles based on the local collision rate in the radial zone. Additionally, both collisions and advection are modeled using the representative particle approach. If the cloud's inner one percent of mass reaches an optical depth of unity during collapse, we set the location where this happens to the accretion radius, the radius at which pebbles are considered accreted onto the optically thick core.

Self-gravity is implemented by using the shell approach: A pebble only feels the gravitational pull from pebbles closer to the center of mass as if the sum of their masses resides as a point source at the center of mass. To go forward in time, the time step for collisions and advection are compared and the time step corresponding to the best resolution is taken to proceed. 

The main conclusions can be summarized as follows:
\begin{enumerate}
    \item Fragmenting collisions in gaseous collapsing pebble clouds with $\mathrm{St_{max}} \sim 0.2$, are rare. The critical terminal velocity $v_{t,*}$ at the location where the cloud becomes optically thick lies far below the fragmentation threshold for which SI will be triggered. Growth through coagulation is negligible, except for the most massive clouds with $R_c \sim 100 $ km and beyond.
    \item Comets and planetesimals collapse toward a primordial core for which the collisions have a negligible effect on altering the initial size distribution.
    \item We find that although the initial size distribution is preserved through the collapse, the order of accretion is that the aerodynamically largest pebbles form the inner core and the pebble sizes decrease toward the surface of the formed core. 
    \item The collapse of pebble clouds is self-similar: Fall timescales are the same for the same Stokes number distribution for increasing cloud mass at given orbital distance. 
    \item Massive clouds are optically thick in the inner region already at t = 0, indicating that during the SI phase these regions will already be highly collisional and possibly form an inner solid core. In this phase, the size-sorting combination of gravity and drag will be less efficient, so the inner core of large objects will not be size-sorted, but will represent the initial size distribution. Only closet to the surface should we see the familiar structure of large grains inside and small grains outside.
    \item Comets that have passed the star are stripped from the loosely bound small pebbles at the surface. The increasingly aerodynamically larger pebbles toward the comet core are harder to lift from the surface. We predict that the core is exposed for these comets with centimeter sized pebbles on the active surface layer. We also predict that comets that have not encountered the star still have the loosely bound millimeter-sized pebble surface layer.
    
\end{enumerate}
Our findings support the notion that comets are primordial in nature and that comets have a systematic radial structure resulting from aerodynamic sorting of the primary building blocks. Future observations and interior analysis of a larger sample of comets are vital to test the validity of our model. 
\begin{acknowledgements}
The authors thank the anonymous referee for a constructive report. We thank Marc Brouwers in particular for useful last moment comments. We thank Tom Konijn, Sjoerd van der Heijden, Michiel Min and Anders Johansen for useful discussions. R.V. acknowledges funding from the Dutch Research Council (NWO), project number ALWGO/15-01. J.D. acknowledges funding from the European Research Council (ERC) under the European Unions Horizon 2020 research and innovation programme under grant agreement No. 714769.
\end{acknowledgements}
\bibliographystyle{aa}
\bibliography{mybibfile}
\appendix
\section{Terminal velocity collapse}
\label{ap:termcoll}
In this appendix, we compute the collapse of a pebble cloud in the limiting case in which the radial velocities are always given by the terminal velocity, at which gravitational force and drag force are equal in size. The terminal velocity
of a pebble is then given by:
\begin{equation}
    v_t = -\frac{G M_\mathrm{enc}}{r_p^2} t_s\quad.
    \label{eq:tvel}
\end{equation}

For $\mathrm{St} \ll 1$, the rate of change in time of the radial position $r_p$ of a pebble is then given by:
\begin{equation}
    \frac{\mathrm{d} r_p}{\mathrm{d} t} =-v_t.
\end{equation}
Integrating this expression
\begin{equation}
    \int_{R_0}^{r_p} r_p'^2 dr' = -\int_{t=0}^{t} dt' GM_\mathrm{enc}t_s,
\end{equation}
with $R_0$ the initial release distance of the pebble and $M_\mathrm{enc} \sim \rho_{0} R_0^3$ the enclosed mass at this particular release distance, gives us $r_p$ as a function of time:
\begin{equation}
    r_p(t) = R_0\left (1 - 4\pi \rho_{0} G t_s t\right )^\frac{1}{3}\quad.
    \label{eq:radvstime}
\end{equation}

The time needed for a pebble to fall toward the cloud center $r=0$ is:
\begin{equation}
    t_t = \frac{\Omega_0}{4 \pi \rho_{0} G  \ \mathrm{St}}\quad,
    \label{eq:falltime}
\end{equation} where we have replaced the stopping time by the Stokes number using equation \eqref{eq:stok}. It is interesting to compare this gas-moderated collapse time with the free-fall time that would apply in the case of no friction 

\begin{equation}
    t_{\rm ff} = \sqrt{\frac{3\pi}{32 G \rho_0}} \quad .
\end{equation}

In the presence of gas the timescale of collapse is effectively a slowed down free-fall collapse. The slow-down factor depends on the coupling strength of pebbles to the gas (Stokes number). A comparison of the numerically determined collapse times with Equation \eqref{eq:falltime} is shown in (\Fg{falltime}). The derivative of \eq{falltime} ${\mathrm{d} t_t}/{\mathrm{d} \mathrm{St}} \propto -1/\mathrm{St}^2$ showing that the difference in fall times become increasingly larger for smaller pebble sizes.  In the limit of large Stokes numbers, the gas-moderated fall time appears to become even shorter than the free fall time.  However, by that time, the assumption that a particle could reach the terminal velocity is no longer valid, in effect, velocities are always limited by the free-fall velocity.
\begin{figure}[t]
    \centering
    \includegraphics[width=.48\textwidth]{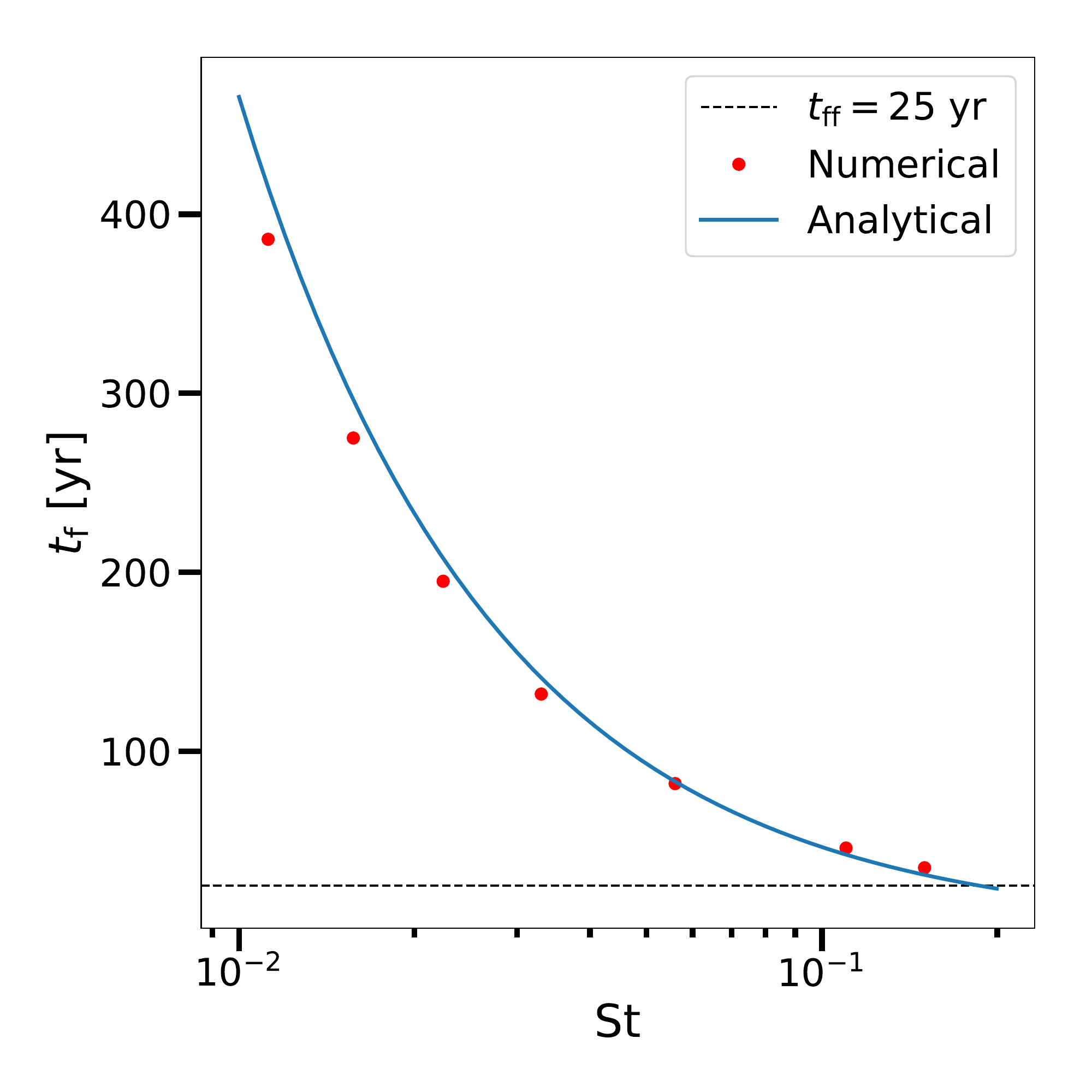} 
    \caption{Fall times in years (y-axis) for pebble clouds composed of a fixed Stokes number (x-axis). The fall times decrease rapidly for increasing Stokes number and the numerical result agrees well with the analytical expectation given in \eq{falltime}.}
    \label{fig:falltime}
\end{figure}
\begin{figure*}[t]
    \centering
    \includegraphics[width=.9\textwidth]{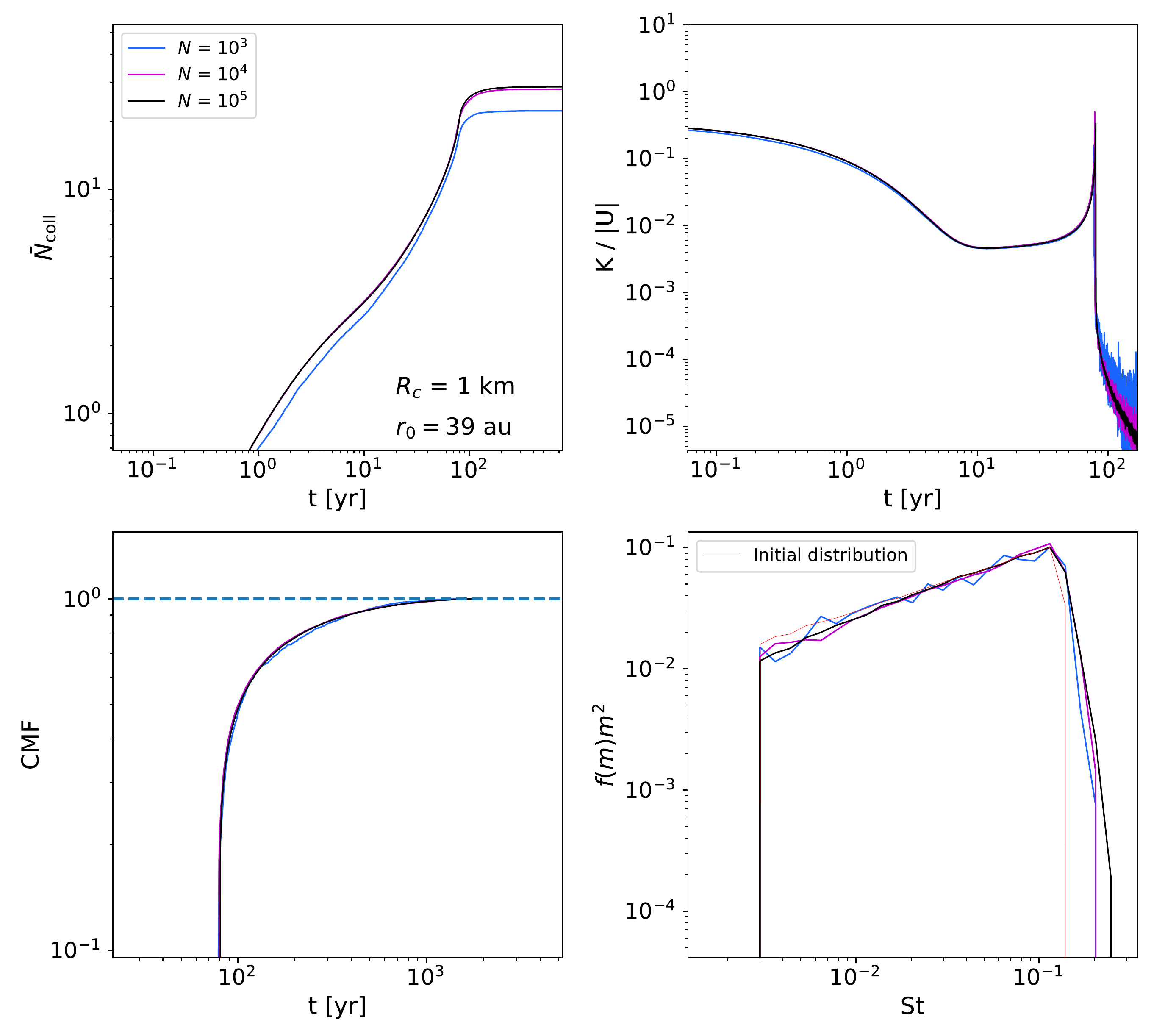}  
    \caption{Convergence test for the fiducial model for different number of representative pebbles $N$. Top left panel: Average collisions per pebble in time. Top right panel: Ratio of the kinetic to potential energy $K/|U|$ of the cloud in time. The results are the same for increasing N already at $N = 10^{4}$ . Bottom left panel: CMF (settled pebbles) in time. Results are the same for all N. Bottom right panel: Mass distribution function vs the radius (Stokes number) of the pebbles of the final core.}
    \label{fig:test1km}
\end{figure*}

\begin{figure*}[t]
    \centering
    \includegraphics[width=.9\textwidth]{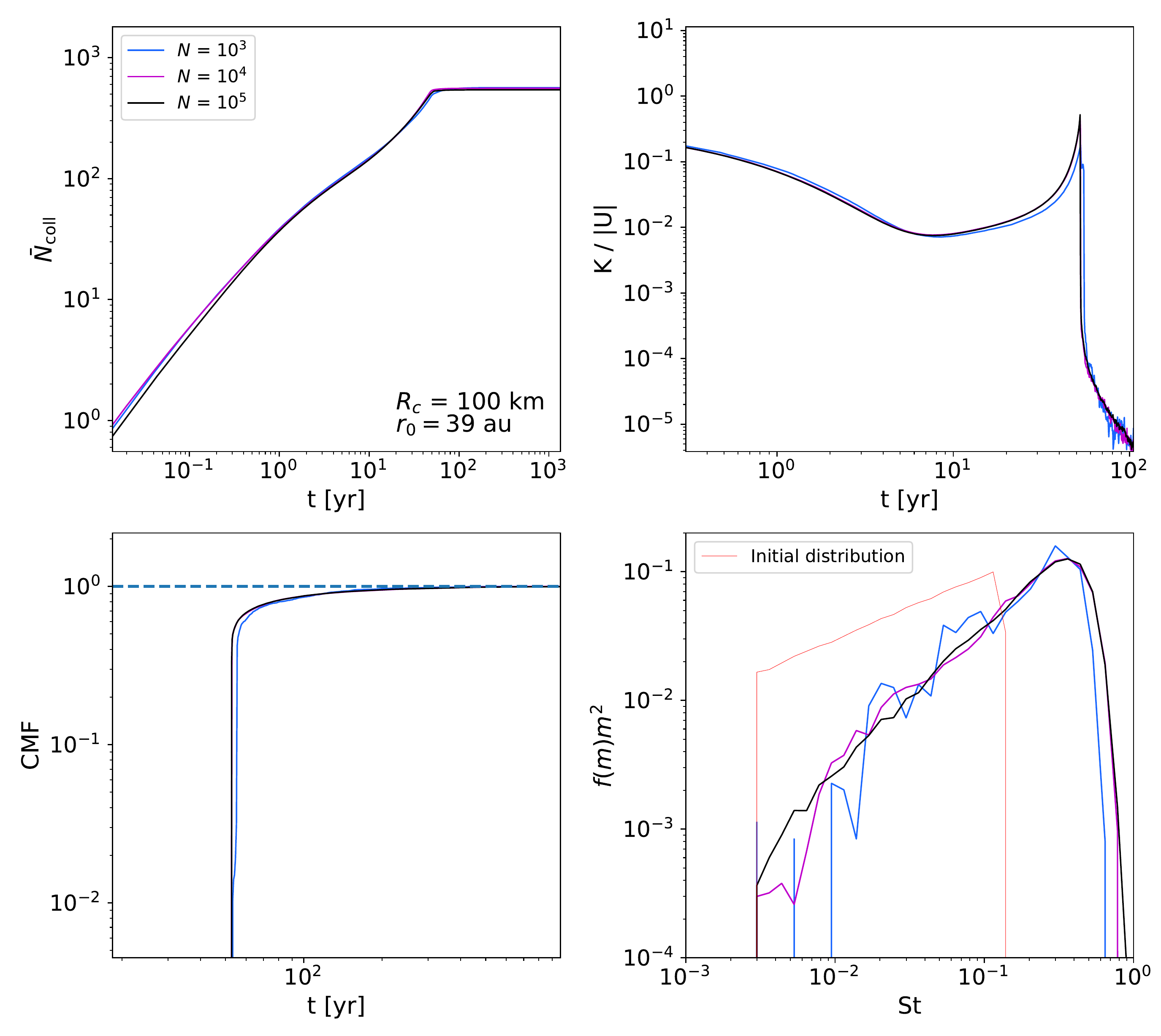}  
    \caption{Same results as discussed in \Fg{test1km} now for 100 km core radius. The most notable difference is a faster collision convergence due to better statistics (more collisions due to higher cloud mass).}
    \label{fig:test100km}
\end{figure*}

\section{Average collisions per pebble}
\label{ap:avcolls}
To determine the importance of collisions, we estimate the total collisions a pebble undergoes before reaching the cloud center. We take an extreme case where the largest pebble $i$ with radius $s_\mathrm{max}$ falls through a column of length $R_\mathrm{H}$ consisting of the smallest pebbles $j$ with radius $s_\mathrm{min}$. We focus on the radial collision rate fed by the terminal velocities of the pebbles. The sweep-out column (average collisions) of the pebble with $s_\mathrm{max}$ before settling is then given by: 
\begin{equation}
    \bar{N}_{\mathrm{coll},i} = n_j \sigma_i R_\mathrm{H}\quad,
    \label{eq:collrate}
\end{equation}
with $n_j = f \rho_0 / m_{p,j}$ the number density of the small pebbles representing a fraction $f$ of the cloud mass, $\sigma_i = 4\pi s^2_i$ the cross-section of the largest pebble. The cross-section of the smallest pebble has been omitted since $s_\mathrm{min} \ll s_\mathrm{max}$. Writing this expression in more global parameters gives:

\begin{align}
\bar{N}_\mathrm{coll,i} \sim 28 \ & \left ( \frac{s_\mathrm{min}}{0.01 \ \mathrm{cm}} \right )^{-3} \left ( \frac{s_\mathrm{max}}{0.5 \ \mathrm{cm}} \right )^{2} \left ( \frac{r_0}{39 \ \mathrm{au}} \right )^{-2} \left ( \frac{\rho_\bullet}{1 \ \mathrm{g \ cm^{-3}}} \right )^{-2/3}\nonumber\times\\
     & \left ( \frac{M_\star}{M_{\odot} } \right )^{2/3} \left ( \frac{R_\mathrm{c}}{1 \ \mathrm{km}} \right )\quad,
    \label{eq:Ncollglobal}
\end{align}

where we have used that the internal density is the same for the pebble and the solid core with radius $R_c$ and the fraction of mass $f$ the small pebbles fill in is estimated at approximately $10^{-2}$. For the used disk radii and core radii in our model the average collisions for the largest pebble are as follows:
At 39 au for 1 km, 10 km and 100 km give $\bar{N}_{\mathrm{coll},i} \sim$ 28, 280, and 2800, respectively, with numerical values from the simulations at $\sim$ 10, 90 and 900, respectively. For the 10 au case, 1 km, 10 km and 100 km give $\bar{N}_{\mathrm{coll},i} \sim$ 53, 530, and 5300, respectively, with numerical values at $\sim$ 20, 100 and 1100. The values agree reasonably well. We note that the optical depth and consequently the accretion radius lower our collision numbers.

\section{Convergence tests and core formation}
\label{ap:convergence}
To verify that results are independent of resolution, we run our model additionally for $N=10^3$ and $N=10^5$ to identify the trend of increasing the resolution. In \Fg{test1km}, we show that $N = 10^4$ provides sufficient resolution to ensure convergence of the results. For the fiducial model, average collisions per pebble $\bar{N}$ (\Fg{test1km}, top left panel) are quite low, which lead to statistically noticable fluctuations. Nevertheless, the average amount of collisions per pebble is reasonably the same for different $N$. For the energy evolution, the graininess increases in the energy curves for lower $N$ since a settled pebble removes a higher contribution from the initial energy of the cloud, both in potential and kinetic energy. The final mass distribution for the core is very similar for all different $N$ (\Fg{test1km}, bottom right panel) indicating that collisional dynamics remains unchanged for sufficient resolution. 

Additionally, we show that for the $R_c = 100$ km case, the convergence in resolution is also ensured in \Fg{test100km}. Specifically, since there are more collisions per pebble, $\bar{N}$ converges sooner in resolution due to better statistics. We focus next on the core formation procedure in our cloud collapse model. 

Since no initial core is present, we quantify core formation by considering the optical depth:
\begin{equation}
    \tau_\mathrm{in}=\sum_{i=1}^{0.01M_t}\frac{\sum_{j=1}^{N_z} N_{j,p} \sigma_j}{4\pi r_i},
\end{equation}
where $N_{j,p}$ is the total amount of physical pebbles one representative particle with mass $m_p$ represents, $N_z$ the amount of representative pebbles in a zone and $\sigma_j$ the cross-section of the physical pebble. $0.01M_t$ represents the total amount of pebbles from the inside out containing the $0.01M_t$ cloud mass and $r_i$ the radial distance of the center of a zone from the cloud center $r = 0$. If $\tau_\mathrm{in} > 1$ a pebble further out cannot penetrate the corresponding layer anymore. The initial core consists of roughly one percent of the cloud mass $M_t$ in the largest pebbles from our size range and reaches $\tau_\mathrm{in} \sim 1$ well before the collapse is finished. This makes sense since the largest pebbles have the shortest fall time. In the occurrence of this core formation criteria, the radius for which pebbles are considered accreted $r_\mathrm{acc}$ is set to the location where $\tau_\mathrm{in} = 1$. Simultaneously we settle the corresponding $0.01M_t$ and repeat the process for the next active inner region. 
\section{Radial shell instabilities}
\label{ap:shells}
In Newton's shell theorem we find that pebbles in a spherically symmetric system only feel the gravitational attraction of the mass situated below them with respect to the center of the cloud $r = 0$, as if this mass were in the center. As a first step, we investigate if pebbles are stable to perturbations in the first place. To find out if an outer pebble situated at initial distance $R_{i + 1}$ can catch up with an inner pebble with distance $R_i < R_{i + 1}$, we equate their distance evolution in time given in \eq{radvstime}. We assume that the system below the pebble at $R_i$ collapses according to \eq{falltime}. We know then that the pebble at $R_i$ also obeys this collapse time and has a distance evolution of:
\begin{equation}
    r_i(t) = R_i\left (1 - 4\pi \rho_{i} G t_s t \right )^\frac{1}{3}\quad.
\end{equation}
The pebble at $R_{i + 1}$ evolves according to:
\begin{equation}
    r_{i + 1}(t) = R_{i + 1}\left (1 - 4\pi \rho_{i + 1} G t_s t \right )^\frac{1}{3} \quad ,
\end{equation}
assuming an equal stopping time. Equating these expressions gives the time after which the two pebbles meet:
\begin{equation}
    t_\mathrm{t,i,i+1} = \frac{R_{i + 1}^3 - R_i^3}{4 \pi G t_s (R_{i + 1}^3 \rho_{i + 1} - R_i^3 \rho_i)}.
    \label{eq:falltime10}
\end{equation}
Using $\rho_{i + 1} = M_{i + 1} / (4\pi R_{i + 1}^3 /3)$ and $\rho_i = M_i / (4\pi R_i^3 /3)$, it becomes
\begin{equation}
    t_\mathrm{t,i,i+1} = \frac{R_{i+1}^3 - R_i^3}{3 G t_s (M_{i+1} - M_i)}\quad.
    \label{eq:masstime}
\end{equation}

Since we are using the enclosed mass approach, we can write $M_{i+1} = M_i + m_s$ where $m_s = M_t / N$ is the mass of one pebble swarm in our numerical simulation. This leads to
\begin{equation}
t_\mathrm{t,i,i+1} = \frac{N(R_{i + 1}^3 - R_i^3)}{3 G t_s M_t}\quad.
\label{eq:masstime2}
\end{equation}

Now we use the spherically uniform radial profile we use for pebbles in the cloud. Pebbles are initiated on positions $r_i = (i / N)^{1/3} R_\mathrm{H}$ with $i$ the pebble number going from 1 to $N$. This means that a pebble put at distance $R_i = R_\mathrm{H}(i / N)^{1/3}$ is followed by a pebble at position $R_{i + 1} = R_\mathrm{H}((i + 1) / N)^{1/3}$ . Filling this in recovers our assumption that pebble $i + 1$ should fall with the same
timescale as the pebble at $R_i$. However, if we place the pebble $i+1$ closer toward the pebble at $i$ at $R_{i + 1} = R_\mathrm{H}(i + f)^{1/3}$, where $f$ is a real number between zero and one and a smaller $f$ puts the ou:ter pebble closer to $R_i$. Plugging this in for $R_{i + 1}$ in \eq{masstime2} gives:
\begin{equation}
    t_\mathrm{t,i,i+1} = \frac{f\Omega_0}{4 \pi G \mathrm{St} \rho_{0}}\quad.
    \label{eq:masstime3}
\end{equation}
This result can be interpreted as follows: The collapse timescale of a pebble with distance $R_{i + 1}$ on a pebble with distance $R_i < R_{i + 1}$ is always shorter than the collapse timescale of the rest of the cloud that is situated below pebble $R_i$. That is, if the distance separation is closer than spherically uniform between the above mentioned pebbles e.a. $0 < f < 1$ (if $f = 0$ the collapse time is zero since they are on top of each other). In the limit of a very high $f$, we see from \eq{masstime3} that the collapse time becomes too high. This makes sense since they will never meet. The expression itself appears to be independent of the resolution $N$. This shows that the radial instabilities occur rapidly if perturbations are triggered, explaining the formation of density peaks during the collapse evolution. The pebble distribution is therefore unstable and minor changes in the initial distribution will lead to regions of high local densities. With a wide size distribution, this does not cause problems.
\begin{table*}[]
\caption{Table of frequently used parameters with a brief description}
\begin{tabular}{ll}
\hline  \hline
\multicolumn{1}{l|}{Parameter}                          & Description                                                  \\ \hline
\multicolumn{1}{l|}{$R_\mathrm{H}$}                     & Hill radius of the core                                      \\
\multicolumn{1}{l|}{$M_t$}                              & Total core mass                                              \\
\multicolumn{1}{l|}{$M_{\star}$}                     & Mass of the star                                             \\
\multicolumn{1}{l|}{$r_0$}                              & Radial distance from the central star                                        \\
\multicolumn{1}{l|}{$f_g$}                              & Gravitational acceleration of the pebble swarm                      \\
\multicolumn{1}{l|}{$M_\mathrm{encl}$}                  & Enclosed mass inside the pebble swarm                                \\
\multicolumn{1}{l|}{$G$}                                & Universal gravity constant                                   \\
\multicolumn{1}{l|}{$r_p$}                              & Distance of pebble swarm from center of mass                 \\
\multicolumn{1}{l|}{$f_d$}                              & Gas drag                                                     \\
\multicolumn{1}{l|}{$t_s$}                              & Stopping time of a pebble                                    \\
\multicolumn{1}{l|}{$v_p$}                              & Speed/velocity of a pebble swarm                             \\
\multicolumn{1}{l|}{$\rho_\bullet$}                     & Internal density of pebbles and core                          \\
\multicolumn{1}{l|}{$s$}                                & Radius of a physical pebble                                  \\
\multicolumn{1}{l|}{$\rho_g$}                           & Gas density                                                  \\
\multicolumn{1}{l|}{$v_\mathrm{th}$}                    & Thermal speed of the gas                                     \\
\multicolumn{1}{l|}{$l_\mathrm{mfp}$} & Mean free path gas molecules                                 \\
\multicolumn{1}{l|}{$\eta_d$}                           & Kinematic viscosity of the gas                               \\
\multicolumn{1}{l|}{$k_b$}                              & Boltzmann constant                                           \\
\multicolumn{1}{l|}{$T$}                                & Gas temperature                                              \\
\multicolumn{1}{l|}{$\bar{m}$}                          & Mean molecular mass                                          \\
\multicolumn{1}{l|}{$H$}                                & Gas scaleheight                                              \\
\multicolumn{1}{l|}{$c_s$}                              & Local sound speed of the gas                                 \\
\multicolumn{1}{l|}{$\Omega_0$}                         & Keplerian frequency                                          \\
\multicolumn{1}{l|}{$\mathrm{St}$}                      & Pebble Stokes number                                         \\
\multicolumn{1}{l|}{$R$}                                & Total collision rate in a zone                               \\
\multicolumn{1}{l|}{$R_{ik}$}                           & Collision rate between pebble swarm $i$ and $k$ in a zone    \\
\multicolumn{1}{l|}{$N_i$}                              & Physical amount of pebbles represented by swarm $i$          \\
\multicolumn{1}{l|}{$\sigma_{ik}$}                      & Cross-section of pebble $i$ and $k$ combined                 \\
\multicolumn{1}{l|}{$\Delta v$}                         & Relative speed between two pebbles                           \\
\multicolumn{1}{l|}{$m_p$}                              & Physical mass of a pebble                                    \\
\multicolumn{1}{l|}{$\mathrm{St}_\mathrm{min/max}$}     & Minimum and maximum Stokes number in simulation respectively \\
\multicolumn{1}{l|}{$U$}                                & Random number between 0 and 1                                \\
\multicolumn{1}{l|}{$\delta t_c$}                       & Time step needed to resolve collisions                       \\
\multicolumn{1}{l|}{$\delta t_\mathrm{EOM}$}            & Timestep needed for advection                                \\
\multicolumn{1}{l|}{$(r,\theta,\phi)$}                  & Spherical coordinates                                        \\
\multicolumn{1}{l|}{$M_\mathrm{swarm}$}                 & Physical mass represented by one pebble swarm                \\
\multicolumn{1}{l|}{$K$}                                & Kinetic energy of the cloud                                  \\
\multicolumn{1}{l|}{$r_\mathrm{acc}$}                   & Accretion radius of the core                                 \\
\multicolumn{1}{l|}{$U$}                                & Potential energy of the cloud   \\
\multicolumn{1}{l|}{$N_\mathrm{sett}$}                  & Total amount of settled pebble swarms                        \\
\multicolumn{1}{l|}{$R_c$}                              & Radius of the core at solid density                          \\
\multicolumn{1}{l|}{tol}                                & Error tolerance advection solver                             \\
\multicolumn{1}{l|}{$\tilde{N}$}                        & Average amount of collisions per pebble swarm                \\
\multicolumn{1}{l|}{$\tau_\mathrm{in}$}                 & Optical depth of inner one percent of cloud mass             \\
\multicolumn{1}{l|}{$v_{t,*}$}                          & Critical terminal velocity for which $\tau_\mathrm{in} > 1$ \\
\multicolumn{1}{l|}{$t_t$}                               & Fall time of a pebble (swarm)             \\                

\end{tabular}

\end{table*}
\end{document}